\documentclass[iop]{emulateapj}

\usepackage{amsmath}
\usepackage{amssymb}

\renewcommand{\vec}[1]{\mathbf{#1}}
\newcommand{\dpar}[2]{\frac{\partial#1}{\partial#2}}
\newcommand{\mstar}{M_{\ast}}
\newcommand{\vgas}{\vec{V}}
\newcommand{\vrgas}{V_{\rm R}}
\newcommand{\vtgas}{V_{\phi}}
\newcommand{\vdust}{\vec{v}}
\newcommand{\vrdust}{v_{\rm R}}
\newcommand{\vtdust}{v_{\rm \phi}}
\newcommand{\sigmagas}{\Sigma_{\rm g}}
\newcommand{\sigmadust}{\Sigma_{\rm d}}
\newcommand{\tstop}{t_{\rm stop}}
\newcommand{\dsize}{s_{\rm d}}
\newcommand{\rhop}{\rho_p}
\newcommand{\Omegak}{\Omega_{\rm K}}
\newcommand{\sonic}{c_s}
\newcommand{\st}{St}
\newcommand{\stp}{St'}
\newcommand{\vk}{V_K}
\newcommand{\vvis}{V_{\rm vis}}
\newcommand{\hg}{h_{\rm g}}
\newcommand{\hd}{h_{\rm d}}
\newcommand{\dgratio}{\sigmadust/\sigmagas}
\newcommand{\mgasdot}{\dot{M}_{\rm g}}
\newcommand{\mdustdot}{\dot{M}_{\rm d}}
\newcommand{\msun}{M_{\odot}}
\newcommand{\vzgas}{V_{z}}
\newcommand{\vzdust}{v_{z}}
\newcommand{\rhogas}{\rho_{\rm g}}
\newcommand{\rhodust}{\rho_{\rm d}}
\newcommand{\dgrhoratio}{\rhodust/\rhogas}
\newcommand{\sch}{Sc}
\newcommand{\sttd}{St_{2D}}
\newcommand{\stptd}{St_{2D}'}
\newcommand{\tstoptd}{t_{\rm stop}^{2D}}

\newcommand{\vrgasavg}{\left< \vrgas \right>}
\newcommand{\vrdustavg}{\left< \vrdust \right>}

\newcommand{\vrgastd}{V_{\rm R}^{2D}}
\newcommand{\vtgastd}{V_{\phi}^{2D}}

\newcommand{\vrdusttd}{v_{\rm R}^{2D}}
\newcommand{\vtdusttd}{v_{\rm \phi}^{2D}}
\newcommand{\vvistd}{V_{\rm vis}^{2D}}
\newcommand{\etatd}{\eta_{2D}}
\newcommand{\sfrag}{s_{\rm frag}}
\newcommand{\vfrag}{v_{\rm frag}}

\newcommand{\au}{\ \rm{AU}}
\newcommand{\yr}{\ \rm{yr}}

\newcommand{\ravg}[1]{\overline{#1}}
\newcommand{\rflc}[1]{#1'}
\newcommand{\ravgrho}{\ravg{\rho}}
\newcommand{\rflcrho}{\rflc{\rho}}
\newcommand{\ravgv}{\ravg{\vec{v}}}
\newcommand{\rflcv}{\rflc{\vec{v}}}

\shorttitle{Effect of dust radial drift on viscous evolution of gaseous disk}
\shortauthors{Kanagawa et al.}

\begin{document}

\title{Effect of dust radial drift on viscous evolution of gaseous disk}

\author{Kazuhiro D. Kanagawa}\email{kazuhiro.kanagawa@usz.edu.pl}
\affiliation{Institute of Physics and CASA$^{\ast}$, Faculty of Mathematics and Physics, University of Szezecin, Wielkopolska 15, PL-70-451 Szczecin, Poland}

\author{Takahiro Ueda}
\affiliation{Department of Earth and Planetary Sciences, Tokyo Institute of Technology, 2-12-1 Ookayama, Meguro, Tokyo, 152-8551, Japan}

\author{Takayuki Muto}
\affiliation{Division of Liberal Arts, Kogakuin University,1-24-2 Nishi-Shinjuku, Shinjuku-ku, Tokyo 163-8677, Japan}

\author{Satoshi Okuzumi}
\affiliation{Department of Earth and Planetary Sciences, Tokyo Institute of Technology, 2-12-1 Ookayama, Meguro, Tokyo, 152-8551, Japan}



\begin{abstract}
The total amount of dust (or ``metallicity") and the dust distribution in protoplanetary disks are crucial for planet formation.
Dust grains radially drift due to gas--dust friction, and the gas is affected by the feedback from dust grains.
We investigate the effects of the feedback from dust grains on the viscous evolution of the gas, taking into account the vertical dust settling.
The feedback from the grains pushes the gas outward.
When the grains are small and the dust-to-gas mass ratio is much smaller than unity, the radial drift velocity is reduced by the feedback effect but the gas still drifts inward.
When the grains are sufficiently large or piled-up, the feedback is so effective that forces the gas flows outward.
Although the dust feedback is affected by dust settling, we found that the 2D approximation reasonably reproduces the vertical averaged flux of gas and dust.
We also performed the 2D two-fluid hydrodynamic simulations to examine the effect of the feedback from the grains on the evolution of the gas disk.
We show that when the feedback is effective, the gas flows outward and the gas density at the region within $\sim 10\au$ is significantly depleted.
As a result, the dust-to-gas mass ratio at the inner radii may significantly excess unity, providing the environment where planetesimals are easily formed via, e.g., streaming instability.
We also show that a simplified 1D model well reproduces the results of the 2D two-fluid simulations, which would be useful for future studies.
\end{abstract}




\section{Introduction} \label{sec:intro}
Terrestrial planets are formed by the accumulation of dust grains in protoplanetary disks \citep[e.g.,][]{Youdin_Goodman2005,Tanaka_Himeno_Ida2005,Okuzumi_Tanaka_Kobayashi_Wada2012}.
In core-accretion scenario \citep[e.g.,][]{Mizuno1980,Kanagawa_Fujimoto2013}, the accumulation of dust grains is also critical for giant planet formation.
How dust grains evolve in protoplanetary disk is directly connected with planet formation.
Moreover, the evolution of dust grains may explain ring structures in protoplanetary disk \citep[e.g.,][]{Zhang_Blake_Bergin2015,Okuzumi_Momose_Sirono_Kobayashi_Tanaka2016}, which are observed in several disks \citep[e.g.,][]{ALMA_HLTau2015,Momose2015,Akiyama2015,Nomura_etal2016,Tsukagoshi2016}.
The evolution of gas and dust grains is one of the most important topics both from theoretical and observational point of view.

Many authors have investigated evolution of dust grains in protoplanetary disk using semi-analytical models \citep[e.g.,][]{Youdin_Shu2002,Dullemond_Dominik2005,Tanaka_Himeno_Ida2005,Birnstiel_Klahr_Ercolano2012,Okuzumi_Tanaka_Kobayashi_Wada2012} and using two-fluid hydrodynamic simulations \citep[e.g.,][]{Zhu2012,Pinilla_Birnstiel_Ricci_Dullemond_Uribe_Testi_Natta2012,Dipierro_Price_Laibe_Hirsh_Cerioli_Lodato2015,Picogna_Kley2015,Jin_Li_Isella_Li_Ji2016,Rosotti_Juhasz_Booth_Clarke2016}.
Dust grains radially drift due to gas--dust friction, and disk gas feels feedback from drifting dust grains.
Most of previous studies considered gas--dust friction only for dust grains and ignored the dust feedback on the disk gas, because grains are not large and the amount of them is negligible as compared with gas.
If the dust grains are highly accumulated or the grains grows up, however, the feedback from dust grains may not be negligible \citep[e.g.,][]{Fu2014,Gonzalez_Laibe_Maddison_Pinte_Menard2015a,Taki_Fujimoto_Ida2016,Gonzalez2017,Dipierro_Laibe2017}.

The feedback from the dust grains affects gas structures.
For instance, \cite{Kretke_Lin_Garand_Turner2009} showed that the inner edge of the dead zone is oscillated by the feedback.
\cite{Taki_Fujimoto_Ida2016} found that the pressure bump in gas is deformed by the feedback from the dust grains trapped in the gas bump.
Recently, using two-fluid (gas and dust grains) SPH simulations, \cite{Gonzalez2017} demonstrated that the grains are trapped in the pressure bump formed by the feedback.
In this paper, we examine how the feedback influences the global evolution of the gas disk induced by viscosity.
In Section~\ref{sec:steady_state}, we describe the structures of gas and dust grains in steady state, considering the vertical settling of the dust grains.
In Section~\ref{sec:sim}, we show the results of 2D two-fluid hydrodynamic simulations and a simplified 1D model, and discuss the effects of the dust feedback on the disk evolution and the planet formation in Section~\ref{sec:discussion}.
Section~\ref{sec:summary} contains our summary.

\section{Structure in steady state} \label{sec:steady_state}
\subsection{Velocities of gas and dust in 2D disk}
Here we consider structures of gas and dust grains in steady state.
First we ignore the vertical structures and simply consider a 2D disk.
We use the polar coordinate ($R,\phi$).
We treat only the surface densities of the gas and the dust, $\sigmagas$ and $\sigmadust$, respectively, instead of densities of the gas and dust, $\rhogas$ and $\rhodust$.
\cite{Kretke_Lin_Garand_Turner2009} derived the formulae of velocities of the gas and the dust grains in the 2D disk.
Very recently \cite{Dipierro_Laibe2017} also derived the similar formulae.
From the equations~(14) and (15) of \cite{Kretke_Lin_Garand_Turner2009}, the radial and azimuthal velocities in the case of single-sized dust grains are given by
\begin{align}
&\vrdusttd = - \frac{2\stptd}{\stptd{}^2+1} \frac{\sigmagas}{\sigmagas+\sigmadust} \etatd \vk \nonumber \\
 & \quad \quad \quad \quad \quad \quad \quad \quad \quad + \frac{1}{\stptd{}^2+1} \frac{\sigmagas}{\sigmagas+\sigmadust} \vvistd \label{eq:vrad_dusttd},\\
&\vtdusttd = \vk  +\frac{1}{\stptd{}^2+1} \frac{\sigmagas}{\sigmagas+\sigmadust} \etatd \vk \nonumber \\
 & \quad \quad \quad \quad \quad \quad \quad - \frac{\stptd}{\stptd{}^2 + 1}\frac{\sigmagas}{\sigmagas+\sigmadust} \vvistd \label{eq:vtheta_dusttd},
\end{align}
and those for the disk gas are given by
\begin{align}
& \vrgastd = \frac{2\stptd}{\stptd{}^2+1} \frac{\sigmadust}{\sigmagas+\sigmadust} \etatd \vk \nonumber \\
& \quad \quad \quad \quad \quad \quad \quad + \left( 1 - \frac{1}{\stptd{}^2 + 1} \frac{\sigmadust}{\sigmagas+\sigmadust} \right) \vvistd \label{eq:vrad_gastd},\\
&\vtgastd = \vk +\left( \frac{1}{\stptd{}^2+1} \frac{\sigmadust}{\sigmagas+\sigmadust} -1 \right)\etatd \vk \nonumber \\
& \quad \quad \quad \quad \quad \quad \quad \quad \quad \quad + \frac{\stptd}{\stptd{}^2 + 1}\frac{\sigmagas}{\sigmagas+\sigmadust} \vvistd \label{eq:vtheta_gastd},
\end{align}
where $\stptd=\sigmagas/(\sigmagas+\sigmadust) \sttd$.
The Stokes number of the dust grains in the 2D disk, $\sttd$, is given by
\begin{align}
&\sttd = \tstoptd \Omegak, \label{eq:st}
\end{align}
where $\tstoptd$ is the stopping time of dust grains in the 2D disk, and $\Omegak = \sqrt{G\mstar/R^3}$ is the Keplerian angular velocity at the mid-plane, where $G$ and $\mstar$ is the gravity constant and the mass of the central star, respectively.
The Keplerian rotation velocity at the mid-plane is $\vk=R\Omegak$.
In the Epstein regime, the stopping time is written by \citep{Takeuchi_Artymowicz2001},
\begin{align}
\tstop (R,z)&=\frac{\rhop \dsize}{\sqrt{8/\pi} \rhogas \sonic},
\label{eq:tstop}
\end{align}
where $\rhop$, $\dsize$, and $\sonic$ are the size and the internal density of dust grains, and the sound speed, respectively.
In the 2D disk, using the surface density, we can write the stopping time as
\begin{align}
\tstoptd(R)&=\frac{\pi \rhop \dsize}{2\sigmagas \Omegak}.
\label{eq:tstop2d}
\end{align}
The pressure gradient force is parameterized by
\begin{align}
&\etatd =-\frac{1}{2} \left(\frac{\hg}{R}\right)^2 \left( \frac{d\ln \sigmagas}{d\ln R} + \frac{d\ln \sonic^2}{d\ln R} \right), \label{eq:eta2d}
\end{align}
and $\vvistd$ is the viscous velocity of gas without dust grains which is given by \citep[e.g.,][]{Lynden-Bell_Pringle1974}
\begin{align}
&\vvistd= -\frac{3\nu}{R} \frac{d\ln \left(\nu \sigmagas R^{1/2}\right)}{d\ln R}, \label{eq:vvis2d}
\end{align}
where $\nu$ is the kinematic viscosity of radial diffusion, given by $\nu = \alpha \sonic \hg$ where the $\alpha$-prescription \citep{Shakura_Sunyaev1973} is used.

In the inviscid case ($\nu = 0$ and thus $\vvistd=0$), only the first terms in RHS of equations~~(\ref{eq:vrad_dusttd})--(\ref{eq:vtheta_gastd}) remain, which are the same as the equations~(2.11)--(2.14) of \cite{Nakagawa_Sekiya_Hayashi1986}.
These terms are originated from gas--dust friction.
On the other hand, if the dust surface density is very small ($\sigmadust \rightarrow 0$), the gas velocities in radial and azimuthal directions correspond to $\vrgastd=\vvistd$ and $\vtgastd=(1-\etatd)\vk$ respectively.

We have employed 2D approximation, meaning that the vertical structures of the gas and dust are similar with each other.
The gas and dust disks are assumed to have the same scale height.
However, the scale height of the dust disk can be much smaller than that of the gas disk, if the dust grains are settled in the mid-plane.
In the next subsection, we discuss the gas and dust velocities, taking into account the vertical structure.

\subsection{Gas and dust motions in a 3D disk} \label{subsec:3ddisk}
We now consider the gas and dust motions in a 3D disk.
We use the cylindrical coordinate ($R$,$\phi$,$z$).
The gas and dust velocities are $\vgas=(\vrgas,\vtgas,\vzgas)$ and $\vdust=(\vrdust,\vtdust,\vzdust)$, respectively.
The equations of motions for the dust grains are given by
\begin{align}
&\dpar{\vrdust}{t} + \left( \vdust \cdot \nabla \right) \vrdust - \frac{\vtdust^2}{R} = -\frac{G\mstar}{R^2}- \frac{\vrdust - \vrgas}{\tstop} \label{eq:eom_dust_rad},\\
&\dpar{\vtdust}{t} + \left(\vdust \cdot \nabla \right) \vtdust + \frac{\vrdust \vtdust}{R} = - \frac{\vtdust - \vtgas}{\tstop} \label{eq:eom_dust_theta},\\
&\dpar{\vzdust}{t} + \left(\vdust \cdot \nabla \right)\vzdust = -\frac{G\mstar}{R^3}z-\frac{\vzdust - \vzgas}{\tstop} \label{eq:eom_dust_z},
\end{align}
where we assume axis-symmetric gravitational potential $\Psi=-G\mstar/\sqrt{R^2+z^2}$, adopting $R \sqrt{1+\left(z/R\right)^2} \simeq R$.
The equations of the motion of the gas can be written by 
\begin{align}
&\dpar{\vrgas}{t} + \left(\vgas \cdot \nabla \right) \vrgas - \frac{\vtgas^2}{R} = -\frac{\sonic^2}{\rhogas} \dpar{\rhogas}{R} -\frac{G\mstar}{R^2} \nonumber \\
& \quad \quad \quad \quad \quad \quad \quad \quad \quad \quad \quad \quad + \frac{f_R}{\rhogas} -\frac{\rhodust}{\rhogas} \frac{\vrgas - \vrdust}{\tstop} \label{eq:eom_gas_rad},\\
&\dpar{\vtgas}{t} + \left(\vgas \cdot \nabla \right) \vtgas + \frac{\vrgas \vtgas}{R} = -\frac{f_{\phi}}{\rhogas}- \frac{\rhodust}{\rhogas} \frac{\vtgas - \vtdust}{\tstop} \label{eq:eom_gas_theta},\\
&\dpar{\vzgas}{t} + \left(\vgas \cdot \nabla \right)\vzgas = -\frac{\sonic^2}{\rhogas} \dpar{\rhogas}{z}-\frac{G\mstar}{R^3} z   +\frac{f_z}{\rhogas} \nonumber \\
&\quad \quad \quad \quad \quad \quad \quad \quad \quad \quad \quad \quad \quad \quad \quad -\frac{\rhodust}{\rhogas} \frac{\vzgas - \vzdust}{\tstop} \label{eq:eom_gas_z},
\end{align}
where $f_R$, $f_{\phi}$, and $f_z$ are the viscous forces in radial, azimuthal, and vertical directions, respectively, which are originated by the gas turbulence.
Assuming the axisymmetric structure, we can express the viscous forces as
\begin{align}
& f_R = \frac{2}{R} \dpar{}{R} \left[ \nu_{rr} \rhogas R \left(\dpar{\vrgas}{R} -\frac{1}{3} \nabla \cdot \vgas \right) \right]  \nonumber \\
& \quad \quad + \dpar{}{z} \left[\nu_{rz} \rhogas \left( \dpar{\vzgas}{z} + \dpar{\vrgas}{R} \right) \right] + \frac{\nu_{\phi \phi}}{R} \left[ 2\frac{\vrgas}{R} - \frac{1}{3} \nabla \cdot \vgas \right], \label{eq:fr_vis}
\end{align}
\begin{align}
f_{\phi} &= \frac{1}{R^2} \left[ \dpar{}{R} \left(\nu_{r \phi} \rhogas R^3 \dpar{\Omega_{\rm g}}{R} \right) + \dpar{}{z}\left(\nu_{\phi z} \rhogas R^3 \dpar{\Omega_{\rm g}}{z} \right) \right] \label{eq:ft_vis},\\
f_z&=\frac{1}{R} \dpar{}{R} \left[\nu_{rz} \rhogas R \left(\dpar{\vzgas}{R} + \dpar{\vrgas}{z} \right) \right] \nonumber \\
& \quad \quad \quad \quad \quad \quad \quad \quad +2\dpar{}{z} \left[\nu_{zz} \rhogas \left( \dpar{\vzgas}{z} - \frac{1}{3} \nabla \cdot \vgas \right)\right] \label{eq:fz_vis},
\end{align}
where $\Omega_{\rm g}=\vtgas/R$, and $\nu_{ij}$ indicates the kinetic viscosity associated with the term of $v_i v_j$ of the Reynolds stress.
The efficiency of the turbulence may be different in direction. 
Hence we distinguish each component of $\nu_{ij}$ in equations~(\ref{eq:fr_vis}) -- (\ref{eq:fz_vis}).
In particular, $\nu_{r\phi}$ and $\nu_{\phi z}$ would be important, which are related with transport of the angular momentum due to radial and vertical shear motions, respectively.
For simplicity, however, we just adopt $\nu_{ij} = \nu$ in the following.
The equations of the motions~(\ref{eq:eom_dust_rad}) -- (\ref{eq:eom_gas_z}) are the same as these adopted in \cite{Nakagawa_Sekiya_Hayashi1986}, excepting the terms associated with the gas viscosity.
However, a parameter range (e.g., dust-to-gas mass ratio, Stokes number of the dust grains, etc.) which the basic equations are valid may not be obvious.
We discuss the validity of these equations in section~\ref{subsec:validity_eq}.
 
The continuity equation of the gas is given by
\begin{align}
& \dpar{\rhogas}{t} + \nabla \left(\rhogas \vgas \right) = 0 \label{eq:continuity_gas},
\end{align}
For the dust grains, the continuity equation is expressed as
\begin{align}
& \dpar{\rhodust}{t} - \nabla \left(\rhodust \vdust + \vec{j} \right) = 0 \label{eq:continuity_dust},
\end{align}
where $\vec{j}$ is the mass flux due to the turbulence of the dust grains \citep[e.g.,][]{Cuzzi_Dobrovolskis_Champney1993,Dubrulle_Morfill_Sterzik1995,Youdin_Lithwick2007}.
From the analogy of the molecular diffusion, assuming the axisymmetric structure, we may obtain $\vec{j} = (j_R,j_{\phi},j_z)$ as \citep[e.g.,][]{Takeuchi_Lin2002,Takeuchi_Lin2005}
\begin{align}
j_R &= \rhogas \frac{\nu_{r\phi}}{\sch} \dpar{}{R} \left(\frac{\rhodust}{\rhogas} \right) \label{eq:j_dust_turb_r},\\
j_{\phi} &= 0 \label{eq:j_dust_turb_th},\\
j_z &= \rhogas \frac{\nu_{\phi z}}{\sch} \dpar{}{z} \left(\frac{\rhodust}{\rhogas} \right) \label{eq:j_dust_turb_z}.
\end{align}

\subsection{Gas and dust structures in steady state} \label{subsec:steady_state3d}
Putting $\partial/\partial t=0$ in the equations of the motion and the continuity equations of the gas and the dust grains described above, we consider the structure of the gas and the dust grains.
The radial and vertical gas velocities can be much smaller than $\vk$ and the deviation of $\vtgas$ from $\vk$ is also small, since the gravity of the central star dominates the gas motion.
We can put $f_r=0$ and $f_z=0$ in equations~(\ref{eq:eom_gas_rad}) and (\ref{eq:eom_gas_z}).

First we consider the vertical structure of the gas and the dust grains in steady state.
Neglecting small advection terms in equations~(\ref{eq:eom_dust_z}) and (\ref{eq:eom_gas_z}), we obtain $\vzdust-\vzgas = -\stp \Omegak z$ and $d\rhogas/dz = -(\rhogas+\rhodust) (\Omegak/\sonic)^2 z$, as shown by \cite{Nakagawa_Sekiya_Hayashi1986}.
Assuming that the vertical gas structure is in hydrostatic equilibrium ($\vzgas=0$), we obtain $\vzdust=-\stp \Omegak z$.
When $\rhogas \gg \rhodust$, moreover, the vertical structure of the gas density can be assumed by
\begin{align}
\rhogas(R,z)&= \rhogas(R,0) \exp\left(-\frac{z^2}{2\hg^2} \right) \label{eq:rhog_vertical},
\end{align}
where $\hg$ is the scale height of the gas disk defined as
\begin{align}
\hg(R)&=\sonic/\Omegak.
\label{eq:gas_scale_height}
\end{align}
Note that when $\rhodust \sim \rhogas$, the thickness of the gas disk may be smaller than $\hg$ given by equation~(\ref{eq:gas_scale_height}) because the dust grains drag the gas as they sediment towards the mid-plane \citep{Nakagawa_Sekiya_Hayashi1986}.
Since equation~(\ref{eq:rhog_vertical}) underestimates the gas density in the mid-plane, we may overestimate the effect of the dust feedback in this case.
For simplicity, however, we use equations~(\ref{eq:rhog_vertical}) and (\ref{eq:gas_scale_height}) even if $\rhodust > \rhogas$.
The validity of this assumption is discussed in section~\ref{subsec:validity_eq}.
Assuming that the radial variations of physical quantities are given by power-law, as $\rhogas(R,0) \propto R^{s}$, $\sonic \propto R^{q/2}$, the angular velocity of gas (without the dust feedback) is described by \citep{Takeuchi_Lin2002}
\begin{align}
	\Omega_{\rm g} (r,z)= \Omegak \left[ 1+ \frac{1}{2} \left(\frac{\hg}{R}\right)^2\left(p + q + \frac{q}{2}\frac{z^2}{\hg^2} \right) \right]. \label{eq:omega_gas}
\end{align}
Hence, the deviation from $\Omegak$ is expressed as $\Omega_{\rm g} = \Omegak \sqrt{1-2\eta}$, where
\begin{align}
\eta(R,z)&=-\frac{1}{2} \left(\frac{\hg}{R}\right)^2 \left(p+q+\frac{q}{2} \frac{z^2}{\hg^2} \right).
\label{eq:eta}
\end{align}
The surface density of the gas is given by
\begin{align}
\sigmagas(R) &= \int^{\infty}_{-\infty} \rhogas dz = \sqrt{2\pi}\rhogas(R,0)\hg(R).
\label{eq:sigmagas}
\end{align}

We obtain the terminal vertical velocity of the dust grains as $-\stp \Omegak z$ \citep{Nakagawa_Sekiya_Hayashi1986}.
The dust grains are settled by the terminal velocity, while the grains diffuse due to the turbulence.
At the steady state, the dust settling is balanced with the turbulence diffusion \citep{Cuzzi_Dobrovolskis_Champney1993,Takeuchi_Lin2002,Takeuchi_Lin2005,Youdin_Lithwick2007}.
In steady state, equation~(\ref{eq:continuity_dust}) gives us
\begin{align}
& \frac{1}{R} \dpar{}{R} \left[ \rhodust \vrdust - \rhogas \frac{\nu_{r\phi}}{\sch} \dpar{}{R}\left(\frac{\rhodust}{\rhogas} \right)\right] \nonumber \\
& \quad \quad \quad \quad + \dpar{}{z} \left[ \rhodust \vzdust - \rhogas \frac{\nu_{\phi z}}{\sch} \dpar{}{z}\left( \frac{\rhodust}{\rhogas} \right) \right] = 0.
\label{eq:continuity_dust2}
\end{align}
Hence, we obtain 
\begin{align}
&\rhodust \vrdust - \rhogas \frac{\nu_{r\phi}}{\sch} \dpar{}{R}\left(\frac{\rhodust}{\rhogas} \right) = F_{m,R}, \label{eq:continuity_dust_rad}\\
& \rhodust \vzdust - \rhogas \frac{\nu_{\phi z}}{\sch} \dpar{}{z}\left( \frac{\rhodust}{\rhogas} \right) = F_{m,z}, \label{eq:continuity_dust_z}
\end{align}
where $F_{m,R}$ and $F_{m,z}$ are mass fluxes of the dust grains in radial and vertical directions, respectively.
Considering the situation that the dust settling is balanced with the diffusion, we put $F_{m,z} = 0$.
Using the terminal velocity of the dust grains, we obtain the vertical distribution of the dust density as \citep{Takeuchi_Lin2002},
\begin{align}
\rhodust(R,z) &= \rhodust(R,0) \exp\left[ -\frac{z^2}{\hg^2} - \frac{\st_{\rm mid}}{\left(\alpha_{\phi z}/\sch \right)} \exp \left( \frac{z^2}{\hg^2} - 1\right) \right] \label{eq:rhod_dist_tl02},
\end{align}
where $\st_{\rm mid}$ is the Stokes number of the dust grains at the mid-plane and $\alpha_{\phi z} = \nu_{\phi z}/(\hg^2 \Omegak)$.
When the dust grains are relatively settled, we expand the expression of $\hd$ with respect to $z/\hg \ll 1$ and take the leading term.
We obtain
\begin{align}
\rhodust(R,z)&= \rhodust(R,0) \exp\left(-\frac{z^2}{2\hd^2} \right) \label{eq:rhod_vertical},
\end{align}
with the scale height of the dust disk given by
\begin{align}
\hd(R)&=\hg(R) \sqrt{\frac{\alpha_{\phi z} /\sch}{\alpha_{\phi z}/\sch+\st_{\rm mid}}}.
\label{eq:ratio_aspectr_gas_dust}
\end{align}
We set $\sch = 1$, because we treat the dust grains with $\st_{\rm mid} < 1$ \citep{Youdin_Lithwick2007}.
The surface density of the dust grains is given by
\begin{align}
\sigmadust(R) &= \int^{\infty}_{-\infty} \rhodust dz = \sqrt{2\pi}\rhodust(R,0)\hd(R).
\label{eq:sigmadust}
\end{align}
Using equations~(\ref{eq:sigmagas}), (\ref{eq:ratio_aspectr_gas_dust}) and (\ref{eq:sigmadust}), we obtain the ratio of $\sigmadust$ to $\sigmagas$ as
\begin{align}
\frac{\sigmadust}{\sigmagas} &= \frac{\rhodust(R,0)}{\rhogas(R,0)} \sqrt{\frac{\alpha}{\alpha+\st_{\rm mid}}}.
\label{eq:dgratio}
\end{align}

Similar to above, we assume $\vrdust,\vzdust$, $\vrgas,\vzgas$, and $\vtdust-\vk$, $\vtgas-\vk$ are much smaller than $\vk$.
Leaving  only the first-order terms with respect to these small values in equations~(\ref{eq:eom_dust_rad}), (\ref{eq:eom_dust_theta}), (\ref{eq:eom_gas_rad}), and (\ref{eq:eom_gas_theta}), we obtain the velocities of the dust grains as
\begin{align}
&\vrdust(R,z) = -\frac{2\stp}{\stp{}^2+1} \frac{\rhogas}{\rhogas+\rhodust} \eta \vk \nonumber \\
& \quad \quad \quad \quad \quad \quad \quad \quad \quad \quad \quad + \frac{1}{\stp{}^2+1} \frac{\rhogas}{\rhogas+\rhodust} \vvis \label{eq:vrad_dust},\\
&\vtdust(R,z) =\vk+\frac{1}{\stp{}^2+1} \frac{\rhogas}{\rhogas+\rhodust} \eta \vk \nonumber \\
& \quad \quad \quad \quad \quad \quad \quad \quad \quad \quad \quad - \frac{\stp}{\stp{}^2 + 1}\frac{\rhogas}{\rhogas+\rhodust} \vvis \label{eq:vtheta_dust},
\end{align}
and those of the gas as
\begin{align}
&\vrgas(R,z) = \frac{2\stp}{\stp{}^2+1} \frac{\rhodust}{\rhogas+\rhodust} \eta \vk \nonumber \\
&  \quad \quad \quad \quad \quad \quad \quad + \left( 1 - \frac{1}{\stp{}^2 + 1} \frac{\rhodust}{\rhogas+\rhodust} \right) \vvis \label{eq:vrad_gas},\\
&\vtgas(R,z) =\vk+ \left( \frac{1}{\stp{}^2+1} \frac{\rhodust}{\rhogas+\rhodust} -1 \right)\eta \vk \nonumber \\
& \quad \quad \quad \quad \quad \quad \quad \quad \quad \quad \quad + \frac{\stp}{\stp{}^2 + 1}\frac{\rhogas}{\rhogas+\rhodust} \vvis \label{eq:vtheta_gas},
\end{align}
where $\stp=\rhogas/(\rhogas+\rhodust) \st$, $\st=\tstop\Omegak$, and $\eta$ is defined by equation~(\ref{eq:eta}).
The viscous velocity of gas (without dust feedback) is given by
\begin{align}
\vvis(R,z)&=-\frac{1}{R} \left[ 3\nu_{r\phi} \frac{d\ln \left(\nu_{r\phi} \rhogas R^{1/2} \right)}{d\ln R}  - q\nu_{\phi z} \frac{d\ln \left(\rhogas z\right)}{d\ln z} \right],
\label{eq:vvis0}
\end{align}
If $\nu_{r\phi}= \nu_{\phi z} = \nu$, we obtain $\vvis$ as
\begin{align}
\vvis&=-\frac{3\nu}{R}\left[ p+ \frac{2q}{3} +2 \left(\frac{z}{\hg}\right)^2 \left(\frac{5q+9}{6}\right) \right].
\label{eq:vvis}
\end{align}

\subsection{Radial net flows of gas and dust grains}
We consider the net mass transfers of gas and dust grains in the disk.
The net radial velocity of the gas are defined by \citep{Takeuchi_Lin2002}
\begin{align}
\vrgasavg(R) &= \frac{1}{\sigmagas} \int^{\infty}_{-\infty} \vrgas \rhogas dz. 
\label{eq:net_vrad_gas}
\end{align}
Similarly, the net radial velocity of the dust grain is defined by
\begin{align}
\vrdustavg(R) &= \frac{1}{\sigmadust} \int^{\infty}_{-\infty} \vrdust \rhodust dz. 
\label{eq:net_vrad_dust}
\end{align}

We adopt the following values as the fiducial case:
$\mstar=1M_{\odot}$, $\rhogas(R=1\au,z=0)=5.3\times 10^{-10} \rm{g/cm}^3$, $\hg=2.8\times 10^{-2} \au$ at $R=1\au$, $p=-2.25$ and $q=-0.5$.
In this case, the gas surface density is obtained by $\Sigma_0 (R/1\au)^{-1}$ with $\Sigma_0=540\rm{g/cm}^2$, and the aspect ratio of the gas disk is proportional to $R^{1/4}$.

\begin{figure}
	\begin{center}
		\resizebox{0.49\textwidth}{!}{\includegraphics{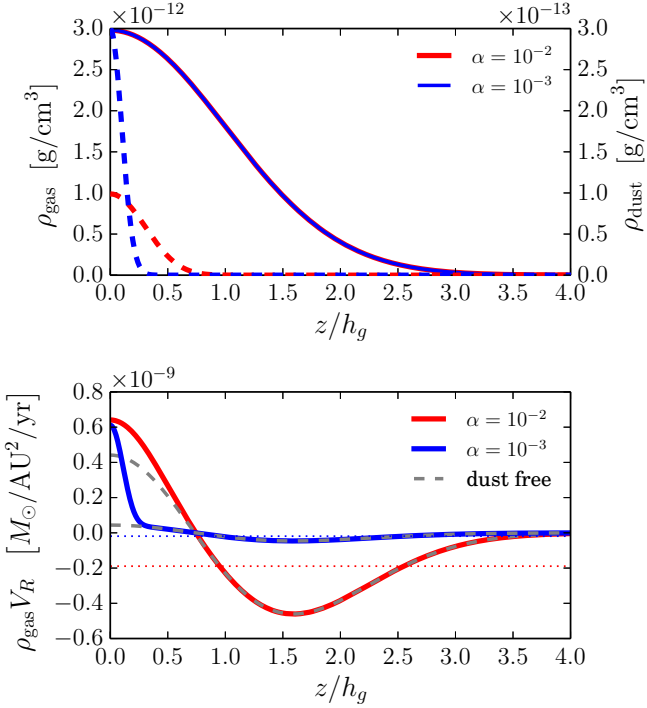}}
		\caption{
		{(\it Top)} The density of gas (solid lines) and dust grains (dotted lines) at $10\au$ ($\rhogas(z=0) = 3.0 \times 10^{-12} \rm{g/cm}^3, \hg/R=0.05$).
		The Stokes number of the dust grain on the mid-plane is set to be $0.1$.
		We set $\sigmadust/\sigmagas=0.01$.
		The color of lines denote the viscosity: red for $\alpha=10^{-2}$, blue for $\alpha=10^{-3}$.
		{(\it Bottom)}
		The vertical distribution of mass flux density of gas at $10\au$.
		The red and blue lines denote $\alpha=10^{-2}$ and $\alpha=10^{-3}$, respectively.
		The grain size is the same as the top panel.
		The gray dashed lines denote the mass flux density without the dust feedback.
		The dotted lines denote the mass flux density in the 2D disk given by $\sigmagas \vvistd/(\sqrt{8\pi}\hg)$.
		\label{fig:vertical_structure}
		}
	\end{center}
\end{figure}
Before presenting the vertical averaged velocities of the gas and the dust , we show typical vertical structures of the gas and the dust grains.
Figure~\ref{fig:vertical_structure} shows the vertical distributions of the gas and dust densities (top) and the mass flux density of gas (bottom) in the cases of $\alpha=10^{-2}$ and $\alpha=10^{-3}$.
In this figure, the Stokes number of the dust grain in the mid-plane is set to be $0.1$, and the ratio of $\sigmadust$ to $\sigmagas$ is $0.01$.
As shown by \cite{Takeuchi_Lin2002}, the gas moves outward near the mid-plane even when the dust feedback is not considered.
However, the dust feedback makes the outward velocity of gas faster.
When $\alpha=10^{-2}$, the gas flows inward above the dust layer ($z \gtrsim \hg$), while the gas in the dust layer flows outward ($z < \hg$).
Above the dust layer, the inward mass flux density is comparable with that assumed in the 2D disk given by $\sigmagas \vvistd/(\sqrt{8\pi} \hg)$.
When $\alpha=10^{-3}$, the thin dust layer is formed near the mid-plane.
The gas moves outward near the mid-plane.
The outward mass flux density near the mid-plane is very large.
On the other hand, above the dust layer, the gas flows inward.
The inward mass flux above the dust layer is the same level as that in the 2D disk, as in the case of $\alpha=10^{-2}$.

\begin{figure}
	\begin{center}
		\resizebox{0.49\textwidth}{!}{\includegraphics{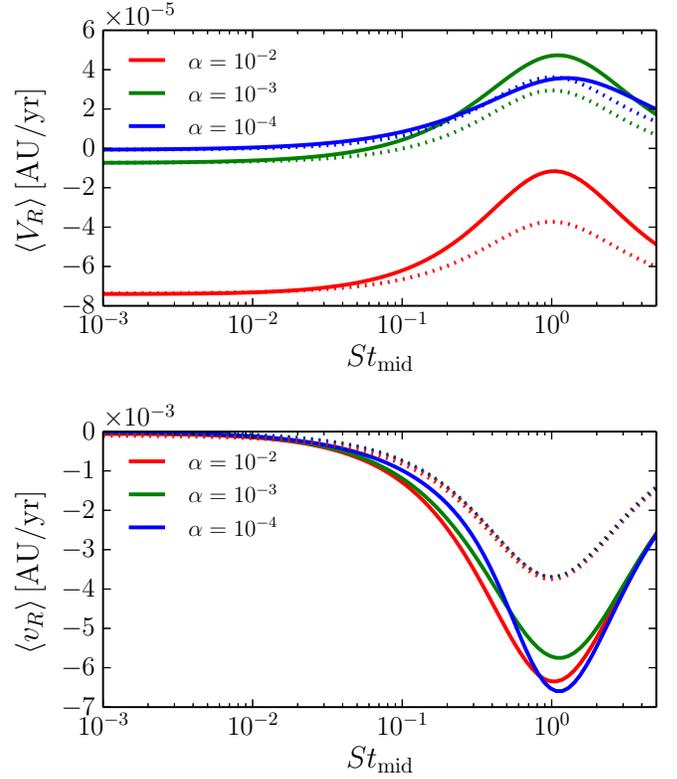}}
		\caption{
		Vertical averaged radial velocities of disk gas (top) and dust grains (bottom) at $R=10\au$ ($\hg/R = 0.05$) and $\dgratio = 0.01$.
		The solid lines are the vertical averaged radial velocity including the dust sedimentation.
		The red, green, and blue line indicate the cases with $\alpha=10^{-2}$, $10^{-3}$, and $10^{-4}$, respectively.
		The dotted lines are the radial velocity of the 2D disk.
		\label{fig:radial_velocities_gas_dust}
		}
	\end{center}
\end{figure}
Figure~\ref{fig:radial_velocities_gas_dust} shows the vertical averaged radial velocities of the gas and the dust grains given by equations~(\ref{eq:net_vrad_gas}) and (\ref{eq:net_vrad_dust}).
For comparison, we also plot the gas and dust radial velocity in 2D disk obtained by equation~(\ref{eq:vrad_gastd}) and (\ref{eq:vrad_dusttd}).
In the figure, we set $\dgratio = 0.01$.
When $\alpha=10^{-2}$, the gas velocity is hardly affected by the grains if the Stokes number is small enough.
As the Stokes number increases, however, the infall gas velocity decreases and when $\st_{\rm mid}=1$, the infall velocity becomes minimum.
As the gas viscosity decreases, the infall velocity due to the viscosity $\vvis$ decreases.
As a result, the radial gas velocity becomes positive due to the feedback from the grains and the gas moves outward.
For instance, when $\alpha=10^{-3}$, a relatively small grain with $\st_{\rm mid}=0.1$ can make the gas flow outward, without any pile-up of grains ($\dgratio=0.01$ in the figure).
For a smaller viscosity (e.g., $\alpha=10^{-4}$), the thickness of the dust layer is very thin.
Although the outward gas flux density in the dust layer is very large, the contribution from the inside of the thin dust layer is not significant.
In fact, the outward velocity when $\alpha=10^{-4}$ is smaller than that when $\alpha=10^{-3}$, since the dust feedback is ineffective in the case of the very small viscosity.
As seen in the bottom panel of Figure~\ref{fig:radial_velocities_gas_dust}, the dust grains always flows inward and the infall velocity is hardly affected by the gas viscosity.

When the size of the dust grain is sufficiently small ($\st_{\rm mid} < 10^{-2}$), the gas and dust radial velocities in the 2D disk given by equation~(\ref{eq:vrad_gastd}) and (\ref{eq:vrad_dusttd}) agree well with the vertically averaged velocities, because the dust layer is not significantly thin.
As the size of grains increases, the velocity in 2D disk deviates from the vertical averaged velocity.
In this case, the dust grains are settled and $\dgrhoratio$ near the mid-plane is much larger than $\dgratio$.
As a result, the effect of the dust feedback is enhanced due to the dust settling for relatively large viscosity such as the cases with $\alpha=10^{-2}$ and $\alpha=10^{-3}$.
On the other hand, if $\alpha$ is very small, the effect of the dust feedback is ineffective because the thickness of the dust layer is very thin.
When $\alpha=10^{-4}$, for instance, the vertical averaged velocity is comparable with that of the 2D disk.
Although the vertical dust settling affects the effect of the dust feedback as discussed above, equation~(\ref{eq:vrad_dusttd}) and (\ref{eq:vrad_gastd}) reasonably reproduce the vertical averaged velocities given by equation~(\ref{eq:net_vrad_dust}) and (\ref{eq:net_vrad_gas}) within a factor of $\sim 2$.

\begin{figure}
	\begin{center}
		\resizebox{0.49\textwidth}{!}{\includegraphics{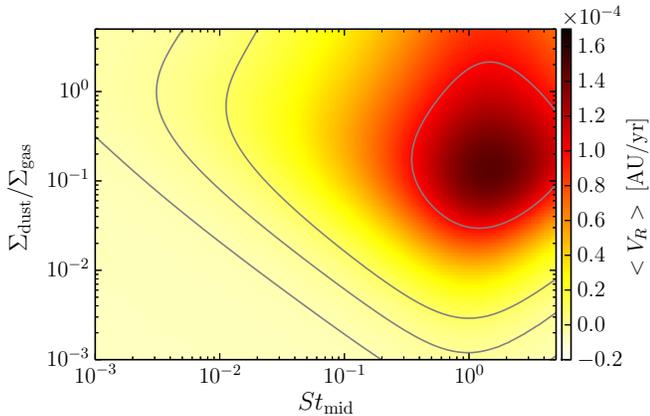}}
		\caption{
		Vertical averaged radial velocity of gas in the case with $\alpha=10^{-3}$ at $R=10\au$ ($\hg/R=0.05$).
		The contour lines show the levels of $\vrgasavg = -5\times 10^{-6} \au/\yr$, $0 \au/\yr$, $10^{-5} \au/\yr$, and $10^{-4} \au/\yr$ from the outside, respectively. 
		\label{fig:vradmap_a1e-3}
		}
	\end{center}
\end{figure}
We show the dependence of $\vrgasavg$ on $\dgratio$ and $\st_{\rm mid}$ when $\alpha=10^{-3}$ in Figure~\ref{fig:vradmap_a1e-3}.
If the Stokes number of the dust grains is sufficiently small, $\vrgasavg$ is negative regardless of the value of $\dgratio$.
As the Stokes number of the dust grain reaches unity, $\vrgasavg$ increases and the gas can move outward for small $\dgratio$.
When $\dgratio$ is large as$\gtrsim 0.5$, $\vrgasavg$ becomes smaller as $\dgratio$ increases.
In this case, $\dgrhoratio$ is much larger than unity at the mid-plane, and hence the gas velocity at the mid-plane becomes small (see, equation~\ref{eq:vrad_gas}).

\begin{figure}
	\begin{center}
		\resizebox{0.49\textwidth}{!}{\includegraphics{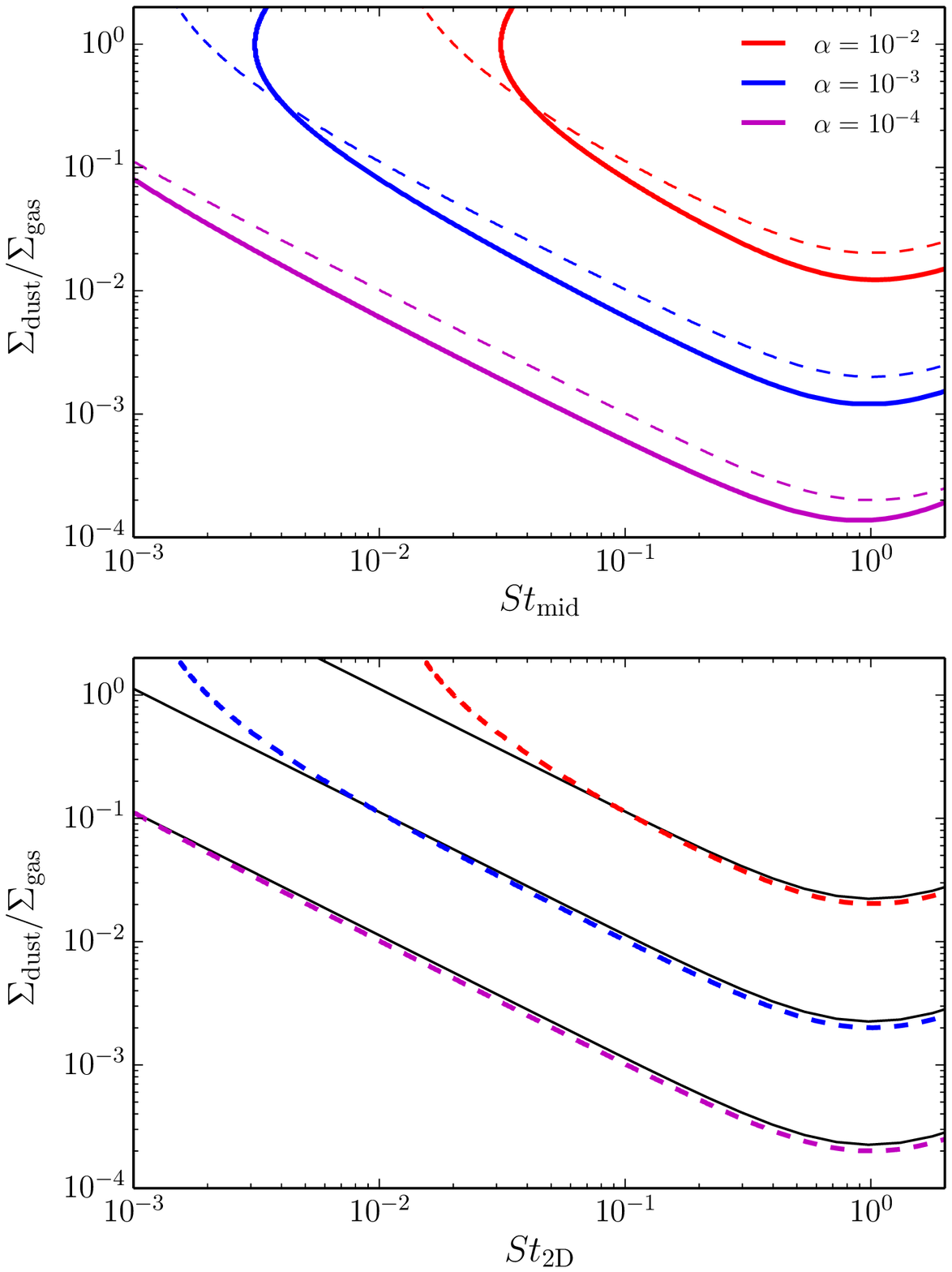}}
		\caption{
		({\it Top}) The dust-to-gas surface density ratio when the vertical averaged radial velocity of the gas (solid lines) is zero at $10\au$, in terms on the Stokes number of the dust grains.
		The dashed lines are the dust-to-gas surface density ratio when $\vrgastd = 0$.
		({\it Bottom})
		The dust-to-gas surface density ratios when $\vrgastd=0$ (dashed lines) and given by equation~(\ref{eq:crit_dgratio}) (solid thin lines).
		\label{fig:dgratio_vrad_zero}
		}
	\end{center}
\end{figure}
In the top panel of Figure~\ref{fig:dgratio_vrad_zero}, we illustrate the relation between the dust-to-gas surface density ratio and the Stokes number of the dust grains when $\vrgasavg=0$.
If the dust-to-gas mass ratio is larger than this critical value, the gas velocity is positive.
When $\alpha=10^{-4}$, the small dust grains with $\st_{\rm mid} = 0.01$ can make the gas move outward if $\sigmadust/\sigmagas \gtrsim 0.01$.
For large dust grains with $\st_{\rm mid} \simeq 1$, the gas can flow outward when the dust-to-gas surface density ratio is only $\sim 10^{-4}$ and $\alpha=10^{-4}$.
Even for the relatively large viscosity, the gas moves outward if relatively large dust grains are highly accumulated (e.g., when $\alpha=10^{-2}$, $\sigmadust/\sigmagas \sim 0.1$ for dust grains with $\st_{\rm mid} \simeq 0.1$).
If the size of the grains is small enough (e.g., $\st_{\rm mid}\simeq 0.05$ in the case of $\alpha=10^{-2}$), on the other hand, the gas moves inward independent of the dust-to-gas surface density ratio.

For comparison, we plot the dust-to-gas surface density when $\vrgastd=0$.
As seen from the top panel of Figure~\ref{fig:dgratio_vrad_zero}, though the $\dgratio$ when $\vrgastd = 0$ is slightly larger than that when $\vrgasavg = 0$, they reasonably agree with each other, if $\dgratio$ is smaller than unity.
Setting $\vrgastd=0$ and assuming $\sigmadust \ll \sigmagas$, we obtain the condition from equation~(\ref{eq:vrad_dusttd}) as
\begin{align}
&\left(\frac{\sigmadust}{\sigmagas}\right)_c \simeq \left( 1+\sttd^{-2} \right)\left( 1+ \frac{2}{\sttd}\frac{\etatd \vk}{\vvistd} \right)^{-1}. 
\label{eq:crit_dgratio}
\end{align}
As seen in the bottom of Figure~\ref{fig:dgratio_vrad_zero}, equation~(\ref{eq:crit_dgratio}) can reproduce the dust-to-gas surface density ratio when $\vrgastd = 0$, if $\dgratio \lesssim 1$.
Equation~(\ref{eq:crit_dgratio}) also reasonably agrees with the dust-to-gas surface density when $\vrgasavg = 0$, if $\dgratio$ is smaller than unity.
Note that the condition of equation~(\ref{eq:crit_dgratio}) corresponds to that the accretion rate of dust grains ($\mdustdot=2\pi R \sigmadust \vrdusttd$) is equal to the gas mass accretion rate due to the viscous diffusion ($2\pi R \sigmagas \vvistd$), in the case of $\sttd \ll 1$.

\subsection{Surface density distributions in gas and dust} \label{subsec:rhosurf_steady_state}
The vertical distribution of the gas flow is affected by the dust settling, which can enhance (and reduce) the effect of the dust feedback, as shown in previous subsections.
However, as seen in Figure~\ref{fig:radial_velocities_gas_dust}, when we consider the net flux integrated over the vertical direction, the approximation of the 2D disk may be reasonable.
In the following, we focus on the net flux of gas and dust, and discuss the evolution of the surface densities.

In steady state, the distribution of the gas surface density is given by $\mgasdot=2\pi R \sigmagas \vrgas = \rm{constant}$ \citep{Pringle1981}.
The gas surface density is given by $\mgasdot/(2\pi R \vrgastd)$.
For a distribution of grains, $\sigmadust = \mdustdot/(2\pi R \vrdusttd)$ \cite[e.g.,][]{Testi_PPVI}.
When $\dgratio \ll 1$ and $\sttd \ll 1$, the radial velocity of dust grains is $\vrdusttd=-2\sttd (\dgratio) \etatd \vk$.
For simplicity, we assume single-size dust grains.
With $\sigmagas \propto R^{-s}$, we obtain $\sigmadust \propto R^{-(s+2f+1/2)}$, where $f$ is the flaring index defined by $\hg/R \propto R^{f}$.
For $\vrgastd$, the first term of equation~(\ref{eq:vrad_gastd}) is approximately $2\sttd (\dgratio) \etatd \vk$ and the second term is $\vvistd$.
If the feedback is negligible, $s=2f+1/2$.
The index of the power law of $\sigmagas$ does not change if the feedback is effective, because the first term in $\vrgastd$ has the same dependence on the radius as $\vvistd$.
Hence, in steady state, $\sigmadust \propto R^{-4f-1}$ and $\dgratio \propto R^{-2f-1/2}$.
In reality, the grain sizes are determined by the coagulation and fragmentation, and are not necessarily constant over the disk.
When turbulent fragmentation limits the size, however, larger dust grains dominate the total mass, and therefore the dust surface density is determined by the maximum size of the grains, which only weakly depends on the location ($\dsize \propto 1/\sqrt{R}$) when $s=1,f=1/4$ \citep{Birnstiel_Klahr_Ercolano2012}.
This situation is not very different from the case with the single-size dust grains.

When the viscosity is sufficiently large but $\dgratio$ is not very large, the gas flows inward in the disk.
As discussed above, the slope of $\sigmagas$ is the same as that in the case that the feedback is negligible.
Hence the distribution of $\sigmagas$ does not change so much, as long as the gas is supplied from outside.
Note that in this case, the accretion rate of gas $\mgasdot$ is reduced because the inflow velocity decreases due to the feedback, but $\sigmagas$ is not changed.
On the other hand, when the viscosity is low or $\dgratio$ is sufficiently large as shown in Figure~\ref{fig:dgratio_vrad_zero}, the gas flows outward in the disk.
In this case, the disk structure with $\mgasdot < 0$ is no longer allowed.
Since there is no gas supply from the inside of the disk in most cases, the disk may be depleted from the inside.

\section{Evolution of gas disk} \label{sec:sim}
\subsection{Model description}
\subsubsection{Two-dimensional simulations}
To examine how the feedback from the grains affects viscous evolutions of gaseous disks, we performed 2D ($R,\phi$) hydrodynamic simulations.
We extended publicly available FARGO code \citep{Masset2000} to include the dust grains.
We simulated the evolutions of disk gas and dust grains by solving the two-fluid (gas and dust grains) equations of motion and continuity.
Because simulating the 2D disk, we do not consider the settling of the dust grains in the simulations.
However, as shown in Figure~\ref{fig:radial_velocities_gas_dust}, the velocities of the gas and the dust grains in 2D disk given by equations~(\ref{eq:vrad_dusttd}) -- (\ref{eq:vtheta_gastd}) reasonably agree with the vertical averaged velocity in the 3D disk.
Hence, the approximation of the 2D disk would be reasonably valid in this case.
The computational domain ranges from $4\au$ to $100\au$ from the central star.
The resolution is 512 and 128 cells in radial and azimuthal directions, respectively.

The initial condition of gas surface density is set as $\sigmagas=\Sigma_0 (R/1\au)^{-1}$ with $\Sigma_0=570 \rm{g/cm}^2$.
We adopt a simple locally isothermal equation of state and assume the disk aspect ratio as $\hg/R = H_0 (R/1\au)^{1/4}$ with $H_0=0.028$.
The initial angular velocity of gas is given by $\Omega_K \sqrt{1-\eta}$, and the mass of the central star is assumed by $1\msun$.
The radial velocity of gas is set by $\vvistd$.
The initial surface density of dust grains is 0.01 of gas everywhere over the disk.
The initial angular velocity of dust grains is $\Omega_K$ and the initial radial velocity is zero.
We adopt dust grains with 3~cm in size, corresponding to the Stokes number $\sttd$ of 0.1 at $R=10\au$ for the initial state.
For simplicity, we neglect coagulation and fragmentation of dust grains, and hence the grain size does not change during the simulations.

Since the gas velocity is very sensitive on the value of $\etatd$ and $\dgratio$ especially in the case of small viscosity, numerical instability occurs when small discontinuities of $\eta$ and $\dgratio$ exist at the innermost part of the computational domain.
To avoid this instability, we introduced a "coupling-damping region" at the innermost radii ($4\au<R<6\au$).
In this region the dust feedback on gas is gradually reduced by $\cos(\pi x^2/2 )$, where $x=R_1-R/(R_1-R_0)$ with $R_0=4\au$ and $R_1=6\au$, respectively.
Hence, at the innermost annulus ($R=4\au$), the gas velocity is set to $\vvistd$.
Moreover, in this region, we force all the physical quantities to be azimuthally symmetric by overwriting the quantities with their azimuthally average at every time step \citep[c.f.,][]{Val-Borro_etal2006}.
That is, in the coupling damping zone, $\sigmagas$, $\vrgastd$ and $\vtgastd$ are related towards their azimuthally averaged values as
\begin{align}
	\frac{dX}{dt}&= - \frac{X-X_{\rm avg}}{\tau_{\rm damp}} f(R),
	\label{eq:damping_force}
\end{align}
where $X$ represents $\sigmagas$, $\vrgastd$ and $\vtgastd$, and $X_{\rm avg}$ denotes the azimuthally averaged values of them, and $\tau_{\rm damp}$ is the orbital period at the boundary.
The function $f$ is a parabola of the form $y=x^2$, scaled to be zero at the boundary layer ($R=6\au$), and unity at the opposite edge ($R=4\au$).
For dust grains, the velocity at the innermost boundary is given by equations~(\ref{eq:vrad_dust}) and (\ref{eq:vtheta_dust}), respectively.
The surface densities of gas and dust are set to be $\dot{M}=\rm{constant}$ \citep[c.f.,][]{Zhu2011}.
At the outer boundary, the velocities of gas and dust are given by equation~(\ref{eq:vrad_dust})-(\ref{eq:vtheta_gas}), and the surface densities are fixed at the initial values.
Here the radial distribution of $\dgratio$ is expected to be smooth.
The turbulent radial mass flux of $j_R$ would be negligible in this case.
Hence we ignore the radial diffusion flux of equation~(\ref{eq:j_dust_turb_r}) for simplicity.

\subsubsection{One-dimensional simulations}
For the purpose of confirming the validity and usefulness of our analytic formulas, we also calculated the evolution by the simplified 1D model with our analytic formulas and compared the results to these of the 2D fluid simulations.
In the 1D model, we simultaneously solved the 1D continuity equation for the dust grains $\partial \sigmadust / \partial t = -(1/R)\partial ( R \sigmadust \vrdusttd)/\partial R$ and that for the disk gas $\partial \sigmagas / \partial t = -(1/R)\partial ( R \sigmagas \vrgastd) /\partial R$. 
Here, we used equations~(\ref{eq:vrad_dusttd}) and (\ref{eq:vrad_gastd}) as the radial velocities of the dust grains and the disk gas ($\vrdusttd$ and $\vrgastd$), respectively. 
The initial and boundary conditions are the same as the 2D simulations described above. 
For the coupling damping region, we just cut off the dust feedback to gas in $R < 6 {\rm AU}$.


\subsection{Results}
\begin{figure*}
	\begin{center}
		\resizebox{0.98\textwidth}{!}{\includegraphics{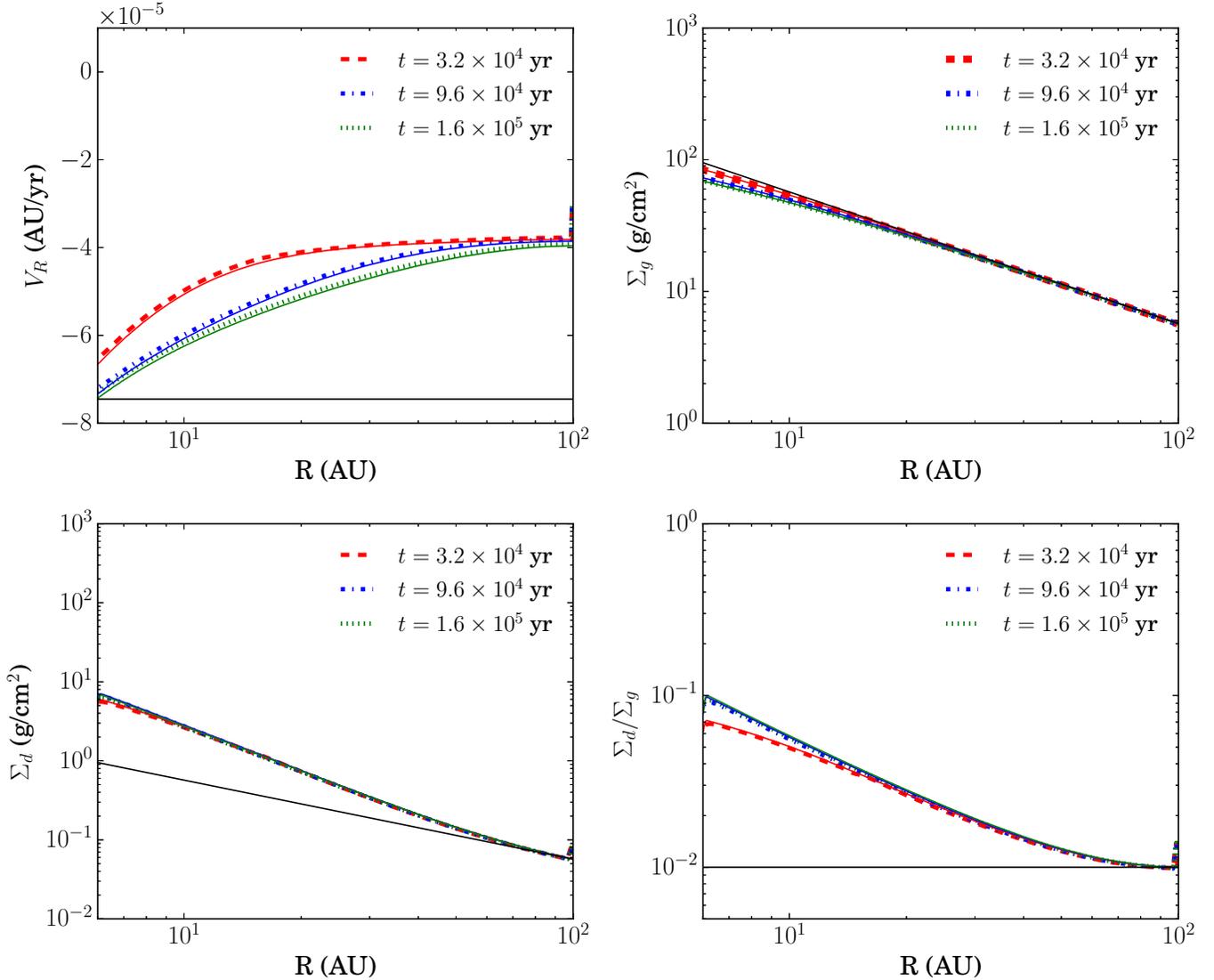}}
		\caption{
		Radial gas velocity (left top), gas surface density (right top), dust surface density (left bottom) rand dust-to-gas surface density ratio (right bottom) in the case with $\alpha=10^{-2}$ and the initial $\dgratio=0.01$.
		The dashed, dot-and-dash, and dot lines represent quantities given by the 2D simulations at $t=3.2\times 10^{4} \yr$, $9.6\times 10^{4} \yr$, and $1.6\times 10^{5}\yr$, respectively.
		The quantities given by 2D simulations are azimuthally averaged in the figure.
		Thin solid lines denote quantities given by the simplified 1D model.
		The solid thin black lines show the initial values.
		\label{fig:evo_a1e-2}
		}
	\end{center}
\end{figure*}
Figure~\ref{fig:evo_a1e-2} illustrates evolutions of the radial velocity and surface densities of gas and dust grains, and dust-to-gas surface density ratio in the viscous case with $\alpha=10^{-2}$.
We compare the results with the 1D model in the figure.
In this case, since the gas viscosity is large and the dust grains is not so concentrated, the gas flows to the central star in the entire region of the disk.
As discussed in subsection~\ref{subsec:rhosurf_steady_state}, the distribution of $\sigmagas$ is hardly changed from the initial distribution which is the steady state without the feedback.
The $\dgratio$ is also distributed as $1/R$, as expected in subsection~\ref{subsec:rhosurf_steady_state}.
Since the velocities of the gas and the dust quickly converge to the values in steady state, the simplified 1D model is able to well reproduce the results of the 2D two-fluid hydrodynamic simulations.

\begin{figure*}
	\begin{center}
		\resizebox{0.98\textwidth}{!}{\includegraphics{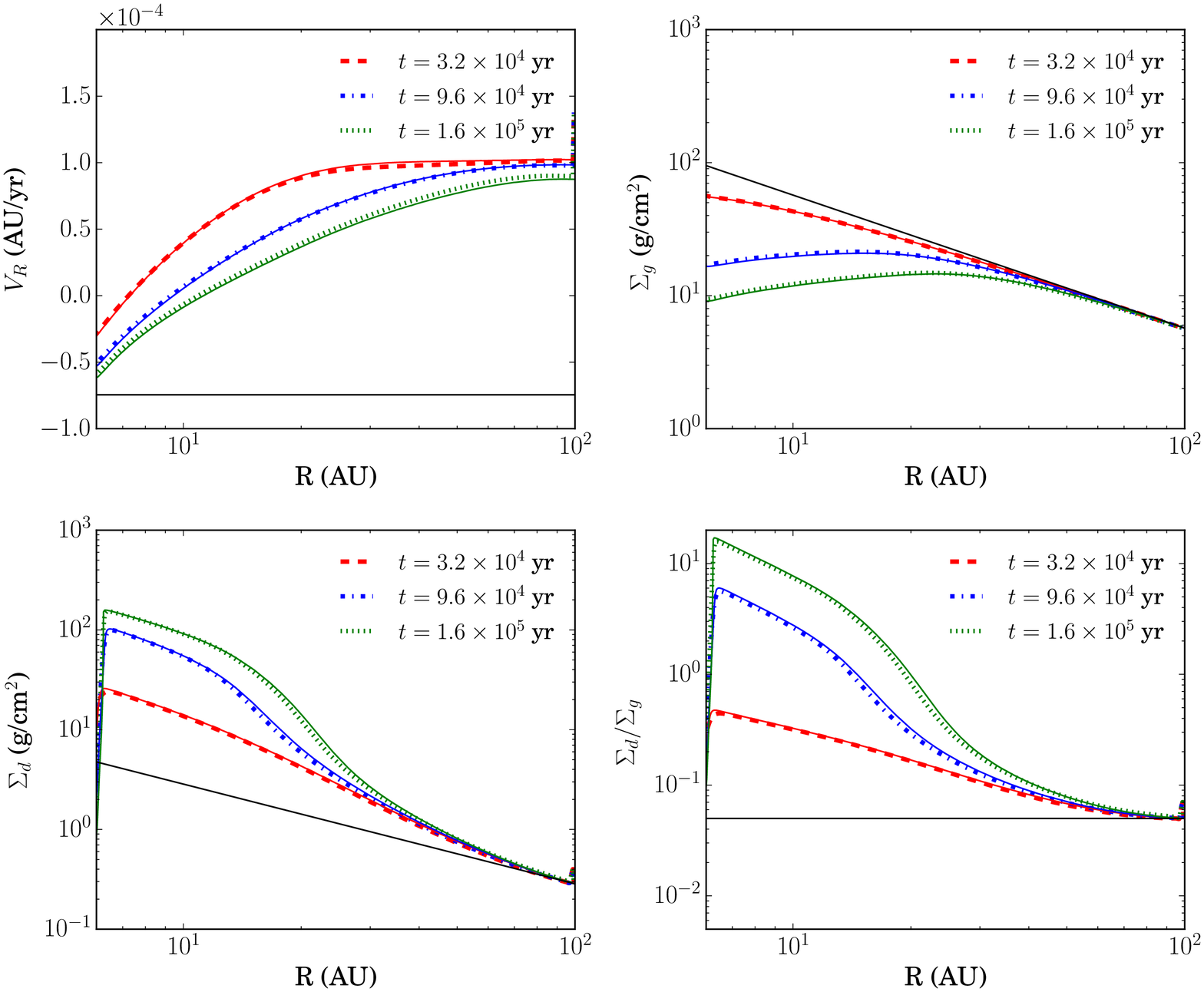}}
		\caption{
		Same as Figure~\ref{fig:evo_a1e-2}, but in the case with initial $\dgratio=0.05$.
		\label{fig:evo_a1e-2_dgr5e-2}
		}
	\end{center}
\end{figure*}

We now show results with different $\alpha$ and $\dgratio$.
The agreement between the 1D model and 2D simulations are always good in the cases presented below.
Figure~\ref{fig:evo_a1e-2_dgr5e-2} shows the evolution of the disk with $\alpha=10^{-2}$ and initial $\dgratio = 0.05$.
Because the initial $\dgratio$ is larger than the case of Figure~\ref{fig:evo_a1e-2}, the gas moves outward in the wide region of the disk.
In this case, the gas surface density decreases in the inner region of the disk.
Owing to the gas removal due to the feedback from the grains, the dust-to-gas mass ratio increases.
The drift velocity of grains decreases as the dust-to-gas mass ratio increases as pointed out by \cite{Drazkowska_Alibert_Moore2016,Ida_Guillot2016} (see also equation~\ref{eq:vrad_dusttd}), which leads to the further increase of the dust-to-gas mass ratio.
Owing to this positive feedback cycle, the inner region of the disk quickly becomes very dust rich.
In this case, after $\sim 10^{5} \yr$, the gas surface density is only $\sim 10\%$ of the initial value at the inner region.
The dust-to-gas mass ratio significantly exceeds over unity, and the Stokes number of grains is $\sim 0.1$.
The dust-to-gas density ratio is also very high.
Using equation~(\ref{eq:dgratio}), we can estimate $\dgrhoratio$ at the mid-plane at $10\au$ as $50$ (at $9.6 \times 10^{4} \yr$) and $150$ (at $1.6 \times 10^{5} \yr$).
Note that the decreases of $\sigmadust$ near the inner boundary ($R=6\au$) is originated from the difference of $\etatd$ in the inside and outside of the coupling damping zone, since $\etatd$ in the damping zone is fixed to the value in the steady state without the feedback.

\begin{figure*}
	\begin{center}
		\resizebox{0.98\textwidth}{!}{\includegraphics{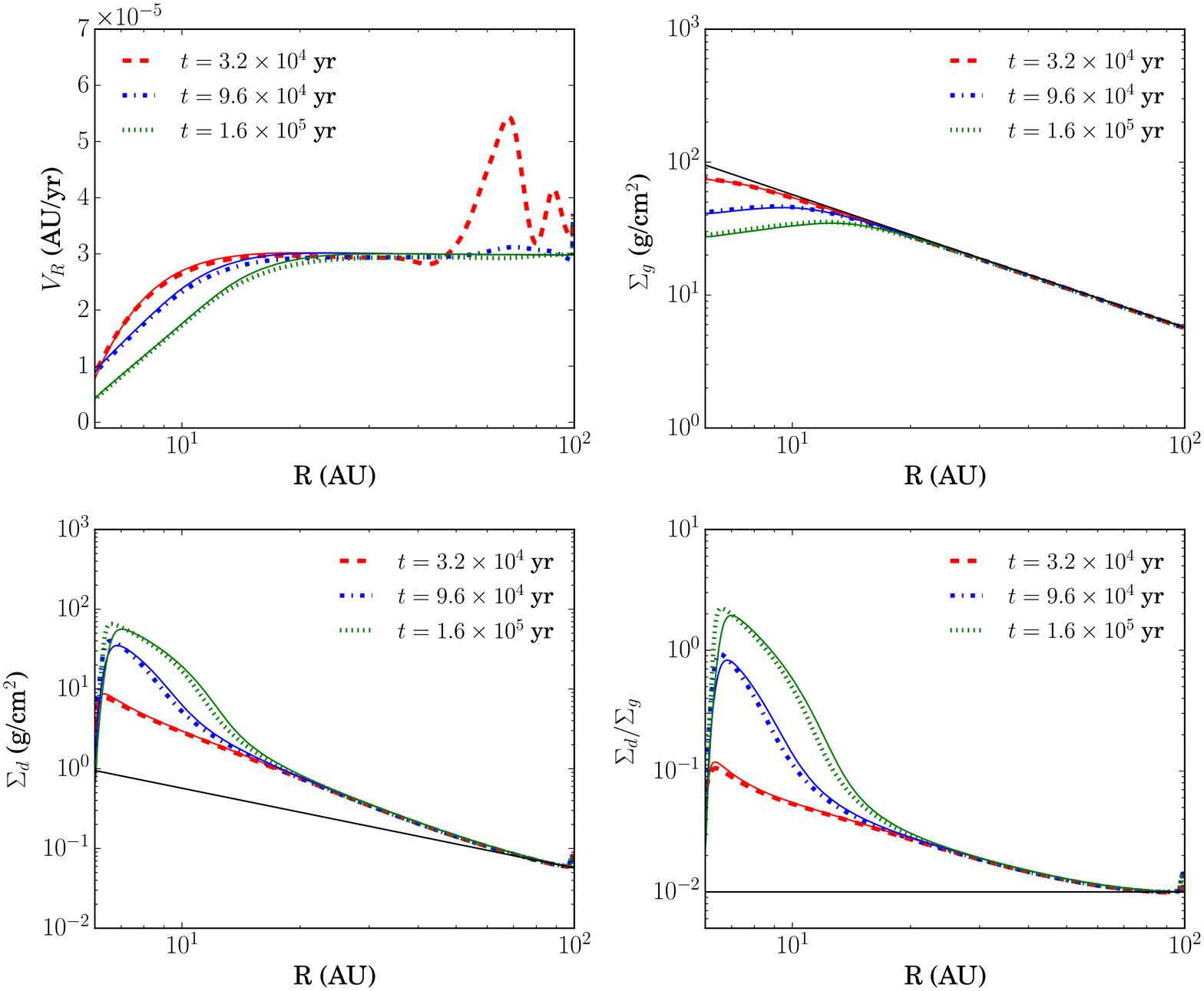}}
		\caption{
		Same as Figure~\ref{fig:evo_a1e-2}, but in the case with $\alpha=10^{-3}$ and the initial $\dgratio=0.01$.
		\label{fig:evo_a1e-3}
		}
	\end{center}
\end{figure*}
The evolution with a lower viscosity ($\alpha=10^{-3}$), is shown in Figure~\ref{fig:evo_a1e-3}.
The initial $\dgratio$ is the same as Figure~\ref{fig:evo_a1e-2}.
Since the viscosity is small, in this case the gas moves outward in the entire disk.
The gas surface density also significantly decreases at the inner region of the disk.
As in the case of Figure~\ref{fig:evo_a1e-2_dgr5e-2}, the removal of gas due to the dust feedback makes the inner region dusty.
After $\sim 10^{5}\yr$, the dust-to-gas surface density ratio increases up to unity.
The dust-to-gas density ratios at the mid-plane at $10\au$ are estimated as $1$ (at $9.6\times 10^{4}\yr$) and $10$ (at $1.6\times 10^{5}\yr$).

\section{Discussion} \label{discussion} \label{sec:discussion}
\subsection{Vertical structure of gas flow}
\begin{figure}
	\begin{center}
		\resizebox{0.49\textwidth}{!}{\includegraphics{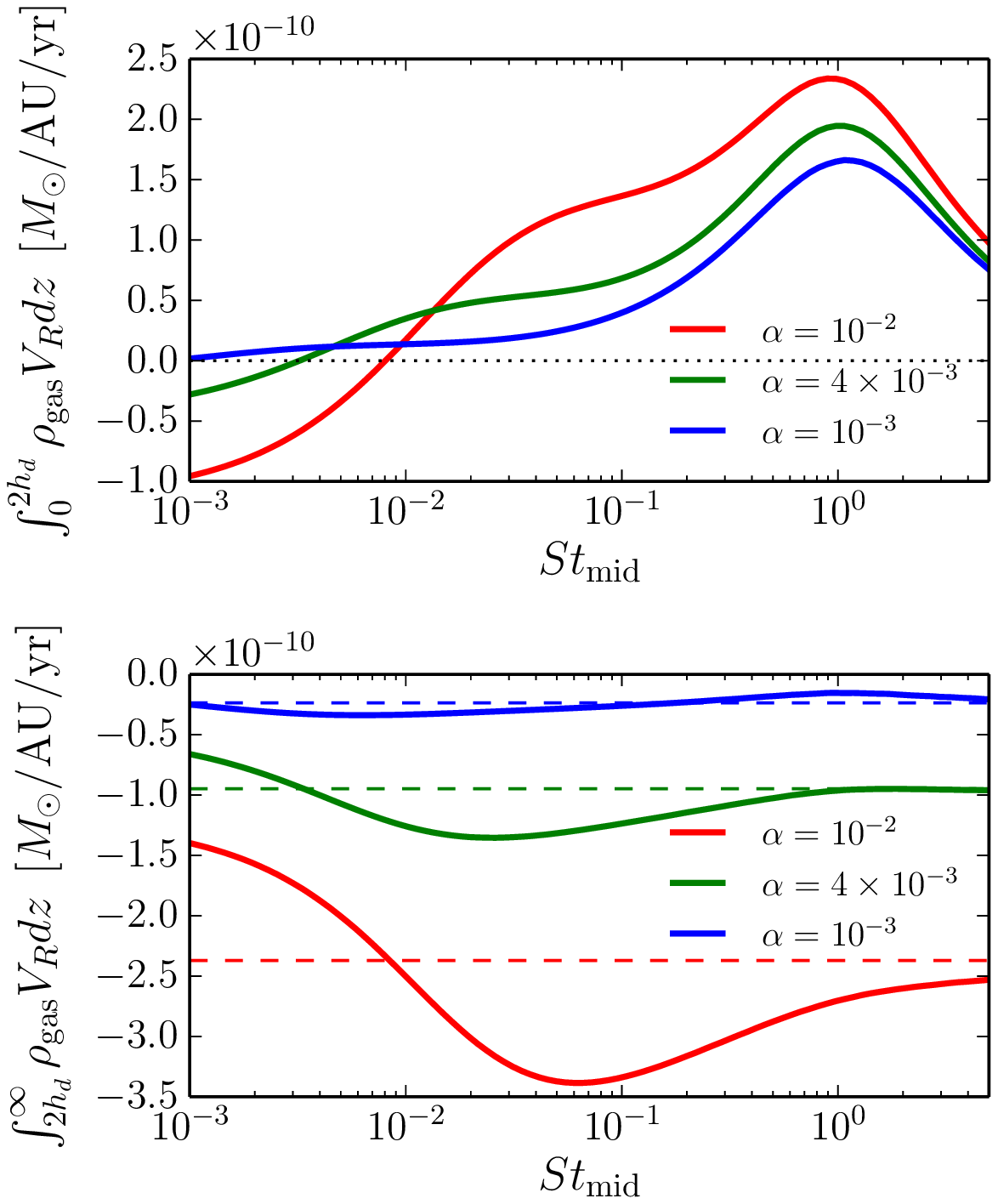}}
		\caption{
		Gas mass fluxes integrated in the inside of the dust layer ($0<z<2\hd$) (top) and the upper of the dust layer ($z>2\hd$) (bottom), when $R=10\au$ and $\dgratio=0.01$.
		The red, green, and blue lines denote the cases with $\alpha=10^{-2}$, $4\times 10^{-3}$, and $10^{-3}$, respectively.
		The dashed thin lines in the bottom panel show the gas mass fluxes without the dust feedback given by $\sigmagas \vvistd/2$.
		\label{fig:inflow_outflow}
		}
	\end{center}
\end{figure}
The gas flows inward when the net radial velocity of the gas $\vrgasavg$ is positive.
As shown in Figure~\ref{fig:vertical_structure}, the mass flux density depends on the altitude.
The dust grains are settled to the mid-plane and the dust layer in which the dust grains are highly concentrated is formed near the mid-plane, while the dust density is quite small above the dust layer.
When the net mass flux is positive, the gas in the upper of the dust layer still flows inward, whereas the gas in the dust layer  flows outward.
In Figure~\ref{fig:inflow_outflow}, we illustrate the mass fluxes integrated in the inside and the upper of the dust layer, in terms of the Stokes number of the dust grains, when $\dgratio=0.01$ at $10\au$ ($\rhogas= 3.0\times 10^{-12}\ \rm{g/cm}^2$, $\hg/R=0.05$).
For a relatively large dust grains (e.g., $\st_{\rm mid} \sim 0.1$), the net radial velocity of the gas is positive.
However, above the dust layer ($z>2\hd$), the gas flows inward.
When $\alpha=10^{-3}$ and $\st_{\rm mid}=1$, for instance, the amount of the gas mass flux above the dust layer is $\sim - 10^{-11} \ \msun/\au/\yr$, which is comparable with the mass flux without the dust feedback($=\sigmagas \vvistd/2$).
This indicates that even when the disk gas deplete from the inside of the disk such as Figures~\ref{fig:evo_a1e-2_dgr5e-2} and \ref{fig:evo_a1e-3}, the gas accretion onto the central star would not stop.
If this disk is observed, we cannot find the depletion of the disk gas, from the viewpoint of the accretion rate onto the central star.
We need to directly observe the disk gas to identify the physical condition of the protoplanetary disk.

\subsection{Dust growth and fragmentation}
In reality, the grain size is strongly limited by the coagulation and fragmentation, though we did not consider this effect in this paper.
However, for silicate grains, since the size is strongly limited by the fragmentation, the size of grains which locally dominates the grain density is not changed so much in the inner region \citep[][]{Birnstiel_Klahr_Ercolano2012}.
In this case, the maximum size of the dust grains $\sfrag$ are given by
\begin{align}
\sfrag &\sim \frac{2}{3\pi} \frac{\sigmagas}{\rhop \alpha} \left(\frac{\vfrag}{\sonic}\right)^2,
\label{eq:sfrag}
\end{align}
where $\vfrag$ is the fragmentation threshold velocity, which depends on the composition (e.g., ice or silicate) of the dust grains.
The fragmentation threshold velocity of the silicate dust grains may be obtained as $1 \rm{m/s}$ -- $10\rm{m/s}$ \citep{Beitz_Guttler_Blum_Meisner_Teiser_Wurm2011}.
When $\alpha=10^{-3}$ and $\sonic = 10^{3} \rm{m/s}$, the Stokes number of the maximum dust grains with $\vfrag = 10 \rm{m/s}$ is about $\sim 0.1$.
In this case, the dust feedback can influence on the entire evolution of the gas disk as shown in Figures~\ref{fig:evo_a1e-2_dgr5e-2} and \ref{fig:evo_a1e-3}, which leads formation of rocky planetesimals. 

If $\vfrag \ll 10 \rm{m/s}$,the dust feedback does not work, since the dust grains cannot grow to a sufficiently large size.
If $\vfrag \gg 10 \rm{m/s}$, the dust grains can grow up to planetesimals without fragmentation, as shown by \cite{Okuzumi_Tanaka_Kobayashi_Wada2012}, or the dust quickly falls to the star because they can quickly grow up to the size of $\st \sim 1$.

\subsection{Implication of planet formation}
When the dust grains are more concentrated as compared with the gas, the grains can grow quickly to planetesimals via the streaming instability \citep[e.g.,][]{Youdin_Goodman2005}.
Roughly speaking, when $\rhodust \gtrsim \rhogas$, the streaming instability may be significantly developed \citep[e.g.,][]{Youdin_Johansen2007,Drazkowska_Dullemond2014,Carrera_Johansen_Davies2015}.
As shown in Figure~\ref{fig:evo_a1e-2_dgr5e-2} and \ref{fig:evo_a1e-3}, when the dust feedback is effective, after $10^{5} \yr$, $\dgratio$ increases over unity within $10\au$.
Using equation~(\ref{eq:dgratio}), we can estimate $\dgrhoratio$ at the mid-plane as $>10$, because $\st \simeq 0.1$.
This ratio is sufficiently high for the streaming instability to occur, though the accurate condition of the streaming instability may depend on the viscosity, the total amount of the dust grains (metalicity) and etc. 
The streaming instability can be caused in more early state of the disk evolution than $10^{5} \yr$.
We should consider the planetesimal formation with the disk evolution.

Once the streaming instability occurs, since the dust grains grow up to the planetesimals, the dust-to-gas mass ratio will decreases.
However, all dust grains are not converted to the planetesimals \citep[e.g.,][]{Johansen_Youdin_Lithwick2012,Simon_Armitage_Li_Youdin2016}.
If the efficiency of the planetesimal formation is sufficiently large to remove the dust grains, the dust feedback will be ineffective.
The gas density will be restored to that in steady state when the dust feedback is not considered.
On the other hand, when the efficiency is moderate, the dust feedback can be still effective after the streaming instability turns on.
In this case, the planetesimals are formed as the gas depletes, which indicates that the migration speed of the planetesimals would significantly decreases.
If the planet is formed, the migration of the planet would also be slow, which is favorable for planet formation.
This situation can continue until most of the dust grains fall to the central star.

\subsection{Implication of disk observation}
Since the disk gas is depleted from the inner region, a hole structure of gas would be formed.
For the dust grains, on the other hand, the density does not decreases in the gas depleted region.
However, if the dust grains are effectively translated to the planetesimals, as discussed in the previous subsection, a density of relatively large grains at the mid-plane which are observed by sub-millimeter may decreases.
In this case, a hole structure can be found by the observations of dust continuum by sub-millimeter, as well as the observation of molecular lines.
Recent ALMA observation done by \cite{Sheehan_Eisner2017} has revealed that the young protoplanetary disk has the hole structure of the dust continuum within $\sim 10\au$.
This hole structure may be explained by the gas depletion due to the dust feedback and the planetesimal formation triggered by the gas depletion.

\subsection{Validity of the model} \label{subsec:validity_eq}
We briefly comment on the validity of the treatment of the dust fluid.
We treat the dust grains as pressureless fluid and ignore some effects of turbulence in the equations of motions (equations~\ref{eq:eom_dust_rad} -- \ref{eq:eom_dust_z}) (see, Appendix~\ref{sec:validty_basic_equations}).
Although this formulation is widely used in the previous studies \citep[e.g.,][]{Fu2014,Gonzalez2017,Dipierro_Laibe2017}, this treatment of the dust fluid may not be appropriate in the cases where the dust density is comparable with the gas density, or the Stokes number of the dust grains is large.
According to \cite{Hersant2009}, the pressure of the dust fluid is not negligible when $\st > 1/2$.
At the mid-plane where the dust density is comparable with the gas density, the dust settling is prevented by the turbulence driven by the instability of the dust fluid \citep[e.g.,][]{Sekiya1998,Sekiya_Ishitsu2000,Johansen_Henning_klahr2006,Chiang2008,Lee_Chiang_Davis_Barranco2010a,Lee_Chiang_Davis_Barranco2010b}.
We must take into account the momentum transfer due to the dust turbulent in this case, and more sophisticated formulation for dust and gas may need to be explored.

Moreover, the vertical structures of the gas and the dust grains may be different from what we  have assumed in equations~(\ref{eq:rhog_vertical}) and (\ref{eq:rhod_vertical}) when the dust density is larger than the gas density.
Gas experiences drag force from the dust grains that sediment towards the mid-plane, so the gas density in the mid-plane is larger than that given by equation~(\ref{eq:rhog_vertical}).
As the dust-to-gas density ratio increases, the settling velocity of the dust grains is slow down ($\vzdust = -\stp \Omegak z \propto 1/(1+\rhodust/\rhogas)$), which leads vertical swelling of the dust layer.
As a result, the vertical distributions of the gas and the dust also deviate from the Gaussian distributions when $\dgratio$ is larger than unity (see, Appendix~A of \cite{Nakagawa_Sekiya_Hayashi1986} or Appendix~\ref{sec:vertical_dens_dist}).
When adopting equations~(\ref{eq:rhog_vertical}) and (\ref{eq:rhod_vertical}), we may therefore have overestimated the dust feedback near a bottom of the dust layer (mid-plane), while underestimating the dust feedback in an upper region of the dust layer.
If considering the vertical averaged quantities, the approximation of the Gaussian distribution may be reasonable.
In any case, if the dust-to-gas density ratio in the mid-plane is of the order of unity and the Stokes number of the dust grains is of the order of unity, the drag force exerted on the gas is at most comparable to the (vertical component of) gravitational force from the central star, and therefore we expect that the qualitative outcomes presented in this paper are not very much affected.
If the dust-to-gas density ratio further increases, it may be necessary to consider more sophisticated formulation of basic equations (e.g., onset of the streaming instability),  which is beyond the scope of this paper.

In summary, our results may be safely used when dust-to-gas mass ratio is less than $\sim 1$, and we expect that qualitative results are valid up to $\dgratio \sim 1$.
If the dust grains are highly concentrated and dominate gas, it may be necessary to have more sophisticated formulation to describe the dynamics of gas and dust.
If streaming instability or other hydrodynamic instabilities may come into play when dust grains dominate, in which case the dust-to-gas mass ratio may not increase too much after all.



\section{Summary} \label{sec:summary}
In this paper, we considered the effect of the dust feedback on the viscous evolution of the gas disk.
Our results are summarized as follows:
\begin{enumerate}
\item We present the analytical expressions for the velocities of gas and dust grains in the 2D disk, considering viscous evolution of gas and gas--dust friction (equations~\ref{eq:vrad_dusttd} -- \ref{eq:vtheta_gastd}).
Considering the vertical dust settling, we also derived the formula of the radial velocities of the gas and dust grains (equation~\ref{eq:net_vrad_gas} with equation~\ref{eq:vrad_gas}).
For the net radial velocities of the gas and the dust, the analytical expression of the 2D disk reasonably agree with the formula considering the dust settling (see Figure~\ref{fig:radial_velocities_gas_dust}).
\item We found that the feedback from the grains significantly affects the radial velocity of the gas (see, Figures~\ref{fig:radial_velocities_gas_dust} and \ref{fig:vradmap_a1e-3}).
When the gas viscosity is sufficiently large and the dust grains are not concentrated, the gas infall slows down due to the dust feedback.
As the viscosity decreases or the dust-to-gas mass ratio increases, the gas flows even outward due to the dust feedback.
\item We also demonstrated the 2D two-fluid hydrodynamic simulations, and showed how the feedback changes the evolutions of gas.
As long as the viscosity is large and the initial dust-to-gas mass ratio is small, the gas flows inward.
In this case, the gas disk evolves as in the case where the dust feedback is not effective, though the infall velocity of the gas decreases (see, Figure~\ref{fig:evo_a1e-2}).
When the viscosity is small or the initial $\dgratio$ is large, the feedback drastically changes the evolution of the disk.
The gas flows to the outside of the disk, and the gas at the inner region is significantly depleted (see, Figures~\ref{fig:evo_a1e-2_dgr5e-2} and \ref{fig:evo_a1e-3}).
The gas removal slows down the infall velocity of the dust grains, and the dust-to-gas mass ratio further increases.
\item We presented the idea of a simplified 1D model.
We solve the continuity equations with the velocities given by equations~(\ref{eq:vrad_dusttd}) and (\ref{eq:vrad_gastd}).
Since the velocities quickly converge to those in the steady state, the simplified 1D model is able to well reproduce the results of the 2D hydrodynamic simulations, as can be seen in Figures~\ref{fig:evo_a1e-2} -- \ref{fig:evo_a1e-3}.
\item We also discuss the effect of the vertical structure on the disk evolution.
If the disk gas is deposited, the gas accretion onto the star is possible.
We need to observe the gas directly to detect the gas depletion.
\item In the inner region of the disk, the size of silicate dust grains would be strongly limited by the fragmentation.
This situation is not very different from the cases assumed in our simulations.
If the silicate grains remove the gas via the feedback, rocky planetesimals would be formed via the streaming instability.
\end{enumerate}
 
In this paper, we pointed out that the dust feedback can significantly influence the evolution of gas disk.
However, we must consider the size evolution of dust grains, because the feedback strongly depends on the grain size.
In order to understand the evolution of gas disk and grains, we should take into account consistently the evolution of grain size and the evolution of gas, with the dust feedback.

\acknowledgments
We would like to thank the anonymous referee for his/her helpful comments.
This work was supported by the Polish National Science Centre MAESTRO grant DEC- 2012/06/A/ST9/00276.
Numerical computations were carried out on the Cray XC30 at the Center for Computational Astrophysics, National Astronomical Observatory of Japan.
S.~O. is supported by Grants-in-Aid for Scientific Research (\#15H02065, 16K17661, 16H04081) from MEXT of Japan.
T.~M. is supported by Grants-in-Aid for Scientific Research (\#26800106, 15H02074, 17H01103) from MEXT of Japan.

\appendix
\section{Validity of basic equations of the gas and dust grains} \label{sec:validty_basic_equations}
In order to describe flows including turbulence, the Reynolds-averaged Navier-Stokes equations of motion are commonly used.
In this way, the density and all velocity components are devided into mean and fluctuating parts as $\rho = \ravg{\rho}+\rflc{\rho}$ and $\vec{v}=\ravgv+\rflcv$ (the former term is the mean part and the later term is the fluctuating part).
The Reynolds-averaged Navier-Stokes equations of the gas and the dust grains in protoplanetary disks are derived by \cite{Cuzzi_Dobrovolskis_Champney1993}.
Here we derive our basic equations~(\ref{eq:eom_dust_rad})--(\ref{eq:eom_gas_z}) from the Reynolds-averaged equations (equations~A.11--A.13 of \cite{Cuzzi_Dobrovolskis_Champney1993}).
As general forms of the Reynolds-averaged equations, \cite{Cuzzi_Dobrovolskis_Champney1993} obtained
\begin{align}
\ravgrho \dpar{\ravg{v_i}}{t} + \dpar{\ravg{\rflcrho \rflc{v_i}}}{t} + \ravgrho \left(\ravg{v_j} \dpar{\ravg{v_i}}{x_j} \right) & = - \dpar{\ravg{P}}{x_i}-\ravgrho \dpar{\Psi}{x_i} + \dpar{\sigma_{ij}}{x_j}\nonumber \\
&-\dpar{}{x_j}\left(\ravg{\rflcrho \rflc{v_j}} \ravg{v_i} \right) - \dpar{}{x_j}\left(\ravg{\rflcrho \rflc{v_i}} \ravg{v_j} \right) +F_i,
\label{eq:RANS_general}
\end{align}
where $P$ and $\Psi$ are a pressure and a gravity potential, respectively, and $\sigma_{ij}$ is the Reynolds stress tensor defined by
\begin{align}
\sigma_{ij} = - \ravgrho \ravg{\rflc{v_i}\rflc{v_j}} - \ravg{\rflcrho \rflc{v_i} \rflc{v_j}},
\label{eq:rstress}
\end{align}
and $F_i$ is the gas--dust friction force given by
\begin{align}
F_i &=\pm \left[\frac{\ravgrho_{d} \left(\ravg{v_{g,i}}-\ravg{v_{d,i}}\right)}{\tstop} + \frac{\ravg{\rflcrho_d \rflc{v_{g,i}}} - \ravg{\rflcrho_d \rflc{v_{d,i}}}}{\tstop} \right],
\label{eq:friction_force}
\end{align}
where the sign of equation~(\ref{eq:friction_force}) is positive for the gas, while it is negative for the dust grains.
As discussed later, if the fluctuating motion of the dust grains is dominated by the gas--dust fraction force, we may neglect the second term of equation~(\ref{eq:friction_force}) (see also, Appendix~B of \cite{Cuzzi_Dobrovolskis_Champney1993}).

First let us consider the equations of the motion for the gas.
If the mach number of the turbulence is not so very high, we can treat the gas fluid as weakly compressible fluid.
In this case, since $\ravgrho_g \gg \rflcrho_g$, we can neglect the time variation of $\ravg{\rflcrho_g \rflc{v_{g,i}}}$ (2nd term of LHS of equation~\ref{eq:RANS_general}) and terms related with the advection due to the turbulence (4th and 5th terms of RHS of equation~\ref{eq:RANS_general}).
Using the Newtonian viscous stress tensor, instead of the Reynolds stress tensor, we obtain the equations of motions for the gas as equations~(\ref{eq:eom_gas_rad}) -- (\ref{eq:eom_gas_z}).

Since we treat the dust fluid as pressureless fluid, $P_d=0$ in equation~(\ref{eq:RANS_general}) for the dust grains.
If adopting the gradient diffusion hypothesis \citep{Cuzzi_Dobrovolskis_Champney1993,Takeuchi_Lin2002}, the mass flux due to the turbulence can be written by
\begin{align}
\ravg{\rflc{\rho_d} \rflc{v_{d,i}}} &= D \ravg{\rho_g} \dpar{\ravg{\rho_d}/\ravg{\rho_g}}{x_i},
\label{eq:massflux_gdy}
\end{align}
where $D$ is the diffusion coefficient which may be given by
\begin{align}
D&=\nu/\sch \label{eq:diffusion_coefficient},
\end{align}
where $\nu$ is the kinetic viscosity of the gas.
In this case, the advection terms due to the turbulence (4th and 5th terms of RHS of equation~\ref{eq:RANS_general}) are proportional to $\nu \ravg{\rho_d}/\sch$.
As long as the distributions of $\ravg{\rho_d}/\ravg{\rho_g}$ and $\ravg{v}$ are not steep, those terms are much smaller than $F_i$ given by equation~(\ref{eq:friction_force}), because $\ravg{\rho_d}\left(\ravg{v_{g,i}}-\ravg{v_{d,i}}\right) \sim \eta \ravg{\rho_d} \vk$.
We can neglect 4th and 5th terms of RHS of equation~(\ref{eq:RANS_general}).

The Reynolds stress of the dust grains may not be negligible.
Here let us consider how the fluctuating motion of the dust grains responses to the fluctuating motion of the gas with the velocity of $v_0$.
Assuming the decay time of the gas fluctuating motion is longer than the time scale that we now consider, we regard that $v_0$ is independent of time.
If the fluctuation motion of the dust grain is dominated by the gas--dust friction, we may write a governing equation of the fluctuation motion of the dust grain as
\begin{align}
\dpar{\rflc{v_{d,i}}}{t} &= -\frac{\rflc{v_{d,i}} - v_0}{\tstop}.
\end{align}
In this case, the fluctuating velocity of the dust grains is given by
\begin{align}
\rflc{v_{d,i}}(t) &= v_0 + \left[\rflc{v_{d,i}}(t=t_0) - v_0 \right] e^{-t/\tstop}. 
\end{align}
The difference between the fluctuating velocities of the gas and the dust grains decays as $\exp(-t/\tstop)$.
For small grains, especially, the velocity of the dust fluctuating motion converges quickly to that of the gas fluctuating motion.
In this case, the correlation $\ravg{\rflc{v_{d,i}}\rflc{v_{d,j}}}$ would be approximately given by $\ravg{\rflc{v_{g,i}}\rflc{v_{g,j}}}$ in equation~(\ref{eq:rstress}).
Similarly, the correlation $\ravg{\rflc{\rho_d}\rflc{v_{d,i}}}$ is also approximated as $\ravg{\rflc{\rho_d} \rflc{v_{g,i}}}$ and they cancel out each other in the second term of equation~(\ref{eq:friction_force}).
When $\ravg{\rho_d} \ll \ravg{\rho_g}$, the components of the Reynolds stress tensor for the dust grains are much smaller than these for the gas becuase the dust density is small.
However, when $\ravg{\rho_d} \sim \ravg{\rho_g}$, the components of the Reynolds stress tensor for the dust grains are comparable with these for the gas.
In this case, we may need to consider the kinetic viscosity of the dust grains in the equation of the motion.
However, since there is a large uncertainty about treatment of the dust fluctuating motion, we drop the term related with the Reynolds stress of the dust grains for simplicity.
Ignoring a time-variation of the turbulence, and the Reynolds stress term and the advection terms due to the turbulence, we derive the equations of motions for the dust grains~(\ref{eq:eom_dust_rad})--(\ref{eq:eom_dust_z}) from the Reynolds-averaged Navier-Stokes equation~(\ref{eq:RANS_general}).


\section{Vertical structures of gas and dust grains} \label{sec:vertical_dens_dist}
The vertical structures of the gas and the dust grains are described by the Gaussian distributions as equations~(\ref{eq:rhog_vertical}) and (\ref{eq:rhod_vertical}).
However, when $\rhodust \gtrsim \rhogas$, the vertical distributions deviate from the Gaussian distributions.

\cite{Nakagawa_Sekiya_Hayashi1986} considered the vertical structure of the gas by adopting a simple dust vertical distribution as $\rhodust(z) = const. = \rhodust{}_{,0}$ for $|z|<\hd$,  otherwise  $\rhodust(z) = 0$ (see Appendix~A of that paper). 
The gas density at the mid-plane is given by $\rhogas(0)\left( 1+ f \dgratio \right)$, where $\rhogas(0)$ is the gas density at $z=0$ given by equation~(\ref{eq:rhog_vertical}), and $f$ is a function of $\hd/\hg$ which is an order of unity (e.g., $f=0.5$ when $\hd/\hg=1$ and $f=1$ when $\hd \ll \hg$).
If $\dgratio \ll 1$, hence, the vertical distribution of gas is given by the Gaussian distribution (equation~\ref{eq:rhog_vertical}).
If $\dgratio \gg 1$, however, the gas density at the mid-plane is larger than that given by equation~(\ref{eq:rhog_vertical}), and the vertical distribution also deviates from the Gaussian distribution.

Here we consider self-consistent vertical structures of the gas and the dust grains.
The vertical gradient of the gas density is given by (see, section~\ref{subsec:steady_state3d}):
\begin{align}
\frac{d\rhogas}{dz} &= -(\rhogas + \rhodust) \frac{z}{\hg^2}\label{eq:zslope_rhog}.
\end{align}
From equation~(\ref{eq:continuity_dust_z}) with $F_{m,z}=0$, the vertical gradient of the dust density is given by
\begin{align}
\frac{d\rhodust}{dz} &= -\rhodust \left[ \frac{\stp}{\alpha/\sch} +1 +\frac{\rhodust}{\rhogas} \right] \frac{z}{\hg^2} \label{eq:zslope_rhod}.
\end{align}
\begin{figure*}
	\begin{center}
		\resizebox{0.98\textwidth}{!}{\includegraphics{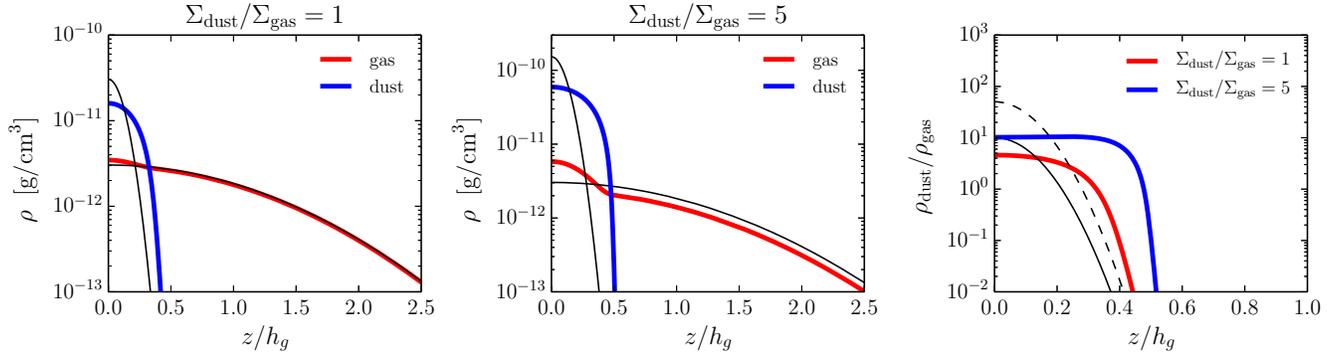}}
		\caption{
		Vertical structures of the gas and the dust grains.
		In the left and middle panels, $\dgratio=1$ and $5$ are adopted, respectively.
		The thin lines are given by the Gaussian distributions of equations~(\ref{eq:rhog_vertical}) and (\ref{eq:rhod_vertical}).
		The distributions of the $\rhodust/\rhogas$ are plotted in the right panel.
		In the right panel, the thin solid and dashed lines are the ratios given by equations~(\ref{eq:rhog_vertical}) and (\ref{eq:rhod_vertical}) for $\dgratio=1$, and $5$, respectively.
		We adopt $\alpha=10^{-3}$, $\st_{\rm mid}=0.1$ and $\sigmagas = 5.7\ \rm{g/cm}^2$.
		\label{fig:vertical_dens_dist}
		}
	\end{center}
\end{figure*}
Solving equations~(\ref{eq:zslope_rhog}) and (\ref{eq:zslope_rhod}), we obtain the vertical structures of the gas and the dust grains.
In Figure~\ref{fig:vertical_dens_dist}, we show the vertical structures of the gas and the dust grains if $\alpha=10^{-3}$ and $\st_{\rm mid} = 0.1$.
For comparison, we plot the Gaussian distributions given by equations~(\ref{eq:rhog_vertical}) and (\ref{eq:rhod_vertical}).
When $\dgratio = 1$, the gas distribution agrees with equation~(\ref{eq:rhog_vertical}), whereas the gas density near the mid-plane is slightly enhanced.
When $\dgratio = 5$, the gas density near the mid-plane increases, while the gas density apart from the mid-plane slightly decreases.
The thickness of the dust layer becomes thicker than that expected by equation~(\ref{eq:rhod_vertical}), because the settling velocity is slow down by the factor of $\rhogas+\rhodust$ in $\stp$.
The dust density at the mid-plane decreases, due to vertical swelling of the dust layer.
The dust-to-gas density ratio is smaller than that expected by the Gaussian distributions of equations~(\ref{eq:rhog_vertical}) and (\ref{eq:rhod_vertical}), while it is larger at the upper part of the dust layer.


\end{document}